\documentclass[aps,pra,reprint,twocolumn,superscriptaddress,showpacs,showkeys,superscriptaddress,longbibliography]{revtex4-1}
\usepackage{amsmath,amssymb,amstext}
\usepackage[usenames,dvipsnames]{color}
\usepackage{graphicx}
\usepackage{bm,bbold,braket}
\usepackage{natbib}
\usepackage{txfonts, comment,stmaryrd}
\usepackage{dcolumn}
\usepackage{color}
\usepackage{xcolor}
\colorlet{rn}{red}
\colorlet{an}{blue}
\usepackage[english]{babel}

\usepackage[colorlinks,bookmarks=false,citecolor=blue,linkcolor=red,urlcolor=blue]{hyperref}
\begin{document}

\title{Landau-Zener transitions and Adiabatic impulse approximation in an array of two Rydberg atoms with time-dependent detuning}
\author{Ankita Niranjan}
\affiliation{Indian Institute of Science Education and Research Pune, 411008, India}
\author{Weibin Li}
\affiliation{School of Physics and Astronomy, University of Nottingham, Nottingham, NG7 2RD, United Kingdom}
\author{Rejish Nath}
\affiliation{Indian Institute of Science Education and Research Pune, 411008, India}
\begin{abstract}
We study the Landau-Zener (LZ) dynamics in a setup of two Rydberg atoms with time-dependent detuning, both linear and periodic, using both the exact numerical calculations as well as the method of adiabatic impulse approximation (AIA). By varying the Rydberg-Rydberg interaction strengths, the system can emulate different three-level LZ models, for instance, bow-tie and triangular LZ models. The LZ dynamics exhibits non-trivial dependence on the initial state, the quench rate, and the interaction strengths. For large interaction strengths, the  dynamics is well captured by AIA. In the end, we analyze the periodically driven case, and AIA reveals a rich phase structure involved in the dynamics. The latter may find applications in quantum state preparation, quantum phase gates, and atom interferometry.
\end{abstract} 

\maketitle

\section{Introduction}
Landau-Zener transition (LZT) between two energy levels occurs when a two-level system is driven across an avoided level crossing. The paradigmatic example being the LZ model in which the diabatic energy levels cross each other linearly in time \cite{lan32, zen32}. The latter had been generalized to both multi-level systems \cite{car86,dem00,mal01,for03,shy04,ban19,ken16, kis13, ash16, sin15,sin16, sin17,fux17,sin17b, par18,mil19} and many-body setups \cite{che10, niu96, wu00, yan09, sal07, kas11, cab12,lar14,zho14}. If driven periodically across an avoided level crossing, the separate LZTs interfere, leading to Landau-Zener-St\"uckelberg (LZS) interferometry \cite{she10}. The LZS interference patterns have been analyzed in various physical setups \cite{par18, she10, van09, shy03, sil06, oli05, wil07, izm08, ste12, dup13, cao13, for14, ota18, liu19}. The interference is attributed to multiple exciting phenomena such as the coherent destruction of tunneling \cite{gro91}, dynamical localization in quantum transport \cite{rag96}, and population trapping \cite{aga94, noe98}. On the application side, the interference features can be utilized to control the qubit states \cite{sai04, gau11, cao13}. 

Different techniques have been employed to analyze the complex dynamics in periodically driven quantum systems \cite{sil17, aba16, pon15, she10}. The most straight forward approach is to solve the corresponding Schr\"odinger equation. Sometimes, specific approximation methods can provide significant insights into the mechanisms involved in quantum dynamics. One successful approach is adiabatic impulse approximation (AIA). While using AIA, the time evolution is discretized into adiabatic and non-adiabatic regimes. It has been employed to study quantum systems undergoing a quench \cite{dam05,dzi12} or periodically driven across an avoided level crossing or a transition point \cite{sil17}. It is thereby analyzing the LZTs and quantum phase transitions, including the Kibble-Zureck mechanism \cite{dam05, dam06, dzi10}. At the impulse point, the transition probability obtained from the LZ model in which the system is driven past the avoided level crossing linearly in time is used \cite{lan32, zen32}.  

Interacting few or many-body periodically driven quantum systems are known to exhibit a variety of new phenomena \cite{dal13,buk15,sil17, aba16, pon15, eck17}. In this regard, Rydberg-excited atoms constitute an ideal platform for such studies \cite{saf10}. Strong interactions between two Rydberg atoms can suppress further Rydberg excitations within a finite volume and is called the Rydberg blockade \cite{luk01,urb09,gae09,hei07}. Rydberg blockade and the breaking of the blockade (anti-blockades) \cite{ate07,qia09,amt10} have been at the heart of the Rydberg based quantum simulators and quantum information applications \cite{saf10}. For two atoms, it has been proposed that through modulation induced resonances, one can engineer the parameter space for both Rydberg-blockade and anti-blockades \cite{bas18}. Periodic modulation in detuning can suppress Rabi couplings, which can lead to selective (state-dependent) population trapping. Not only that, periodic driving in Rydberg gases provides us insights into fundamental problems, but also finds applications in developing robust quantum gates \cite{hua18, wu19}. To implement periodic driving in a Rydberg chain,  one can modulate the light field which couples the ground to the Rydberg state. Another way is to apply additional radio-frequency or microwave fields, and they provide off-resonant couplings to other Rydberg states. The two methods respectively create sidebands either in the driving field or in the atomic levels \cite{aut55, gal08}.  Rydberg atoms in oscillating electric fields \cite{zhe15} have been explored experimentally for manipulating the dipole-dipole interactions via F\"orster resonances \cite{tre14, zhe215, tau08, boh07}. Also, LZTs across a F\"orster resonance is probed in an experiment using a frozen pair of Rydberg atoms in which the dipole-dipole interaction is vital  \cite{saq10}. But most of the experiments probing LZTs are limited to either a single Rydberg excitation or conditions in which the Rydberg-Rydberg interactions (RRIs) are non-relevant \cite{rub81, noe98, rob00, con02, lam02, gur04, mae11, fey15, zha18}.

In this paper, we analyze the dynamics in two two-level atoms in which the ground state is coupled to a Rydberg state with a time-dependent detuning. We consider both linear and periodic variation of detuning in time. Before indulging in the two-atom case, we revisit the AIA for a single two-level atom. The exact results are in an excellent agreement with that from AIA under suitable criteria. Also, we identify a striking similarity between the expression for the excitation probability obtained via AIA for the periodically driven case and the intensity distribution of the narrow, equal-amplitude, multi-slit (or a uniform antenna array) interference pattern. The two-atom setup features three distinct avoided level crossings, and it realizes a bow-tie LZ model for vanishing interactions and a triangular LZ model for strong RRIs. The energy gaps and the energetic separation between the avoided crossings, the two relevant parameters in LZ dynamics, can be modified by varying RRIs. Also, the ratio between the interaction strength and the square root of the quench rate plays a vital role. We observe various features in the LZ dynamics, for instance, Rabi-like oscillations in diabatic states, sharp LZ transitions between adiabatic states at large RRIs, and beats in the triangular LZ model. At large RRIs, AIA captures the exact dynamics accurately.

\label{sa}
\begin{figure}
	\centering
	\includegraphics[width= .95\columnwidth]{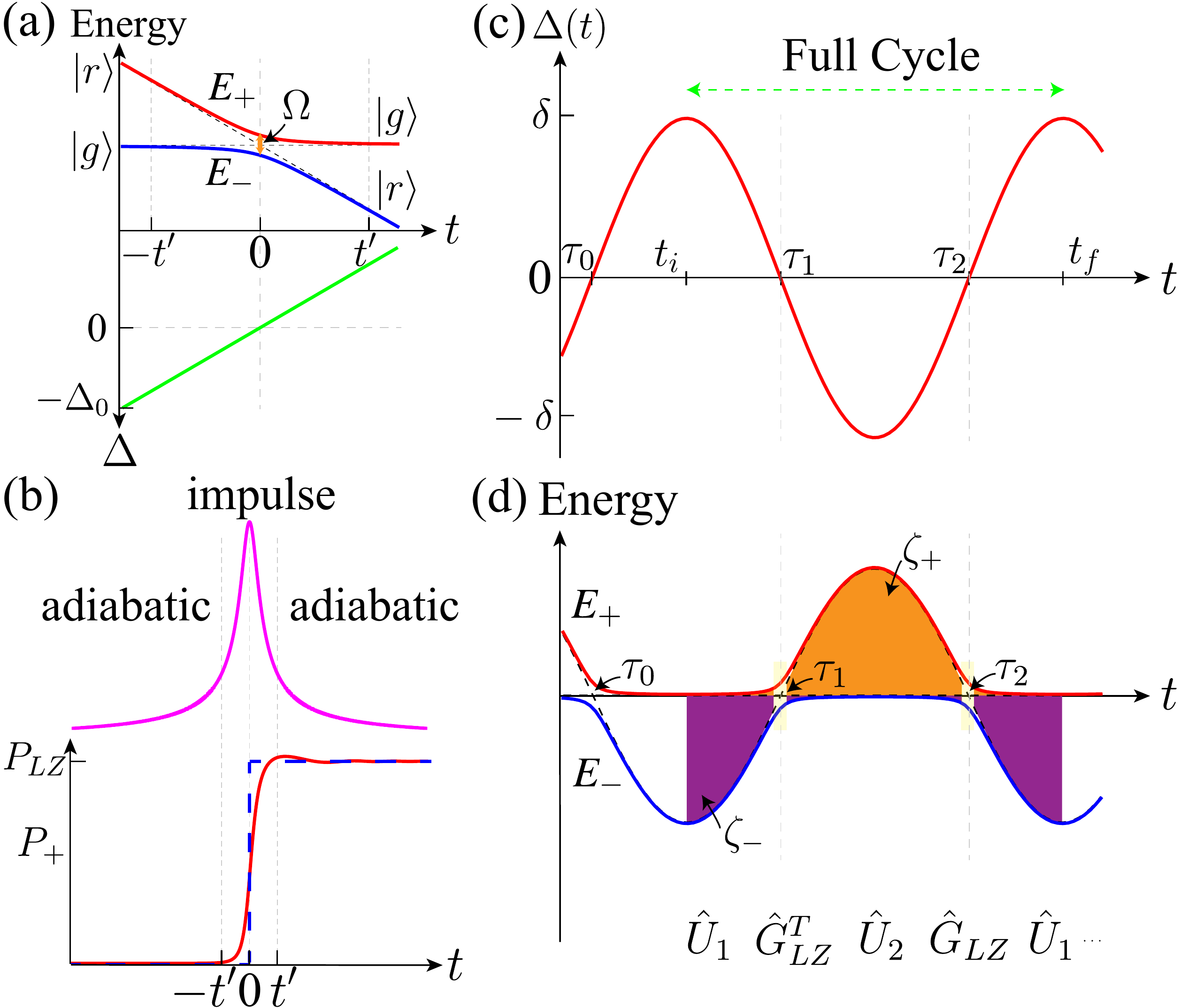}
	\caption{(a) The linear variation of detuning and the instantaneous energy eigenvalues. The dashed lines show the diabatic energy levels. (b) shows the inverse of the energy-gap between the adiabatic levels (top one), and we approximately identify the impulse, and the adiabatic regimes, separated at $\pm t'$. The bottom plot shows the population in the excited state ($P_+=|a_+|^2$) vs time $t$ for which the atom is initially prepared in the ground state and subject to a linear quench in the detuning. Solid line shows the exact result and dashed line is from the LZ model [Eq. (\ref{plz})]. (c) depicts the periodic time dependence of $\Delta(t)$, and (d) shows the corresponding instantaneous energy eigenvalues $E_{\pm}$. At the avoided crossings, $t=\tau_{2n} (\tau_{2n+1})$ LZT takes-place, described by the operator $\hat{G}_{LZ} (\hat{G}_{LZ}^T )$ and either side of it the adiabatic evolution takes place, determined by $\hat U_1$ and $\hat U_2$. The shaded area indicates the accumulated phases ($\zeta_{\pm}$) during the adiabatic evolution.}
\label{fig:1}
\end{figure}

In the end, we look at the case of periodically modulated detuning especially, for large RRIs. The latter assures that the avoided crossings are well isolated, and each of them involves only two adiabatic states. When the detuning is modulated across the first avoided crossing, at shorter periods, the dynamics is found identical to that of a two-level atom. At more extended periods, due to resonances, all the three levels become relevant, resulting in the violation of AIA. As the amplitude of modulation gets larger, incorporating other avoided crossings, more resonances emerge in the dynamics. For AIA to capture the exact dynamics, all the adiabatic states must be involved in the LZTs. Besides that AIA reveal the rich structure of phases involved in the dynamics, including the dynamical ones. The detailed information about phases could be very relevant in applications such as the coherent preparation of quantum states, implementing quantum (phase) gates, and atom-interferometry.

The paper is structured as follows. In Sec.~\ref{s1}, we review the dynamics in a two-level atom subjected to time-dependent detuning, both linear and periodic in time. We introduce the concepts of AIA, and the exact numerical results are compared to that of AIA. The validity criteria for AIA is discussed. In Sec.~\ref{ksl} results from AIA for a periodically driven atom is compared to the multi-slit interference pattern. In Sec.~\ref{2as}, we extend the studies to the two-atom setup. The three-level LZ model is analyzed in Sec. \ref{lzm}. Different cases based on the initial states are considered, and population dynamics in both adiabatic and diabatic basis are discussed, including the formation of beats [see Sec.~\ref{bea}]. In Sec.~\ref{aia2}, the results from exact numerics for the three-level LZ model is compared to that of AIA. Finally, the periodically driven setup is studied in Sec.~\ref{per}, and based on the driving amplitude various cases are studied. We summarize in Sec.~\ref{summ}.

\section{Single Two-level Atom and Adiabatic Impulse Approximation}
\label{s1}
In this section, we briefly summarize the LZ dynamics in a single two-level atom for both linear and periodic variation of detuning. The two-level atom constitutes of the ground state $\ket{g}$ and a Rydberg state $\ket{r}$, driven by a laser field with a Rabi frequency $\Omega$ and a time-dependent detuning $\Delta(t)$. We neglect the motional dynamics of the atom and the system is described by the Hamiltonian ($\hbar=1$), 
\begin{equation}
\hat{H}(t) = \frac{\Omega}{2} \hat{\sigma}_x -\Delta (t) \hat{\sigma}_{rr} ,
\label{h1}
\end{equation}
where $\hat{\sigma}_{rr} = \ket{r}\bra{r}$ and $\hat{\sigma}_x = \ket{g}\bra{r} + \ket{r}\bra{g}$ are projection and transition operators, respectively. The states $\{\ket{g} , \ket{r} \}$ form  the diabatic basis whereas the adiabatic basis consists of the instantaneous eigenstates of the Hamiltonian, $\hat{H}(t) \ket{\phi_\pm (t)} = E_\pm(t)\ket{\phi_\pm(t)}$. The time-dependent energy eigenvalues are $E_\pm (t) = \pm \frac{\Omega}{2}\beta_\mp(t)$ with $\beta_\pm (t) = \left[\bar{\Omega}(t)\pm \Delta(t)\right]/\Omega$ and $\bar{\Omega}(t) = \sqrt{\Delta(t)^2 + \Omega^2}$. The variation of $E_\pm (t)$ with the instantaneous detuning for both the linear and the periodic variation in time is shown in Figs. \ref{fig:1}(a) and \ref{fig:1}(d), respectively. The adiabatic and diabatic bases are related to each other by the time-dependent coefficients $\beta_\pm(t)$ via
\begin{equation}
\ket{\phi_\pm(t)} = \sqrt{\frac{\Omega}{2\bar{\Omega} }}\left( \pm \sqrt{\beta}_\pm \ket{g} + \sqrt{\beta}_\mp  \ket{r} \right),
\end{equation}
Far away from the avoided level crossings ($|\Delta|>>\Omega$), the adiabatic levels converge to the diabatic states [see Fig. \ref{fig:1}(a)]. The exact dynamics of the system is obtained by numerically solving the Schr\"odinger equation: $i\partial/\partial t |\psi(t)\rangle=\hat H|\psi(t)\rangle$. In the adiabatic basis, we can write $|\psi(t)\rangle=a_+(t)\ket{\phi_+(t)}+a_-(t)\ket{\phi_-(t)}$, where $a_\pm(t)$ is the time-dependent probability amplitude for finding the atom in the instantaneous adiabatic states $\ket{\phi_\pm(t)}$.

\subsection{Adiabatic Impulse Approximation}
\label{aiasa}

\begin{figure}
	\centering
	\includegraphics[width= 1.\columnwidth]{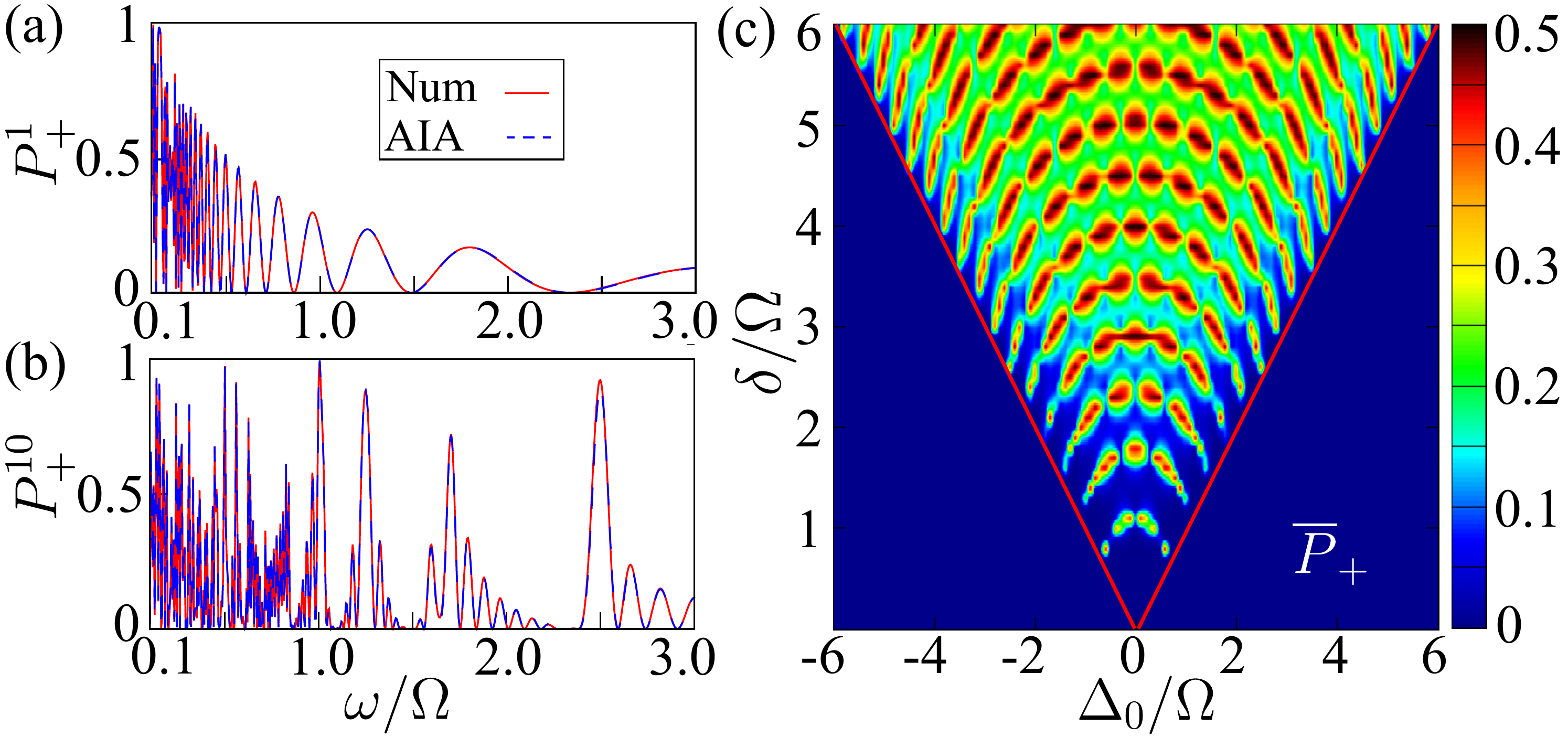}
	\caption{The transition probability to the excited state as a function of $\omega$ when the atom is initially prepared in the ground state for $\delta=20\Omega$, $\Delta_0=5\Omega$, $t_i = \pi/2\omega$ after (a) one cycle  and (b) 10 cycles. The solid line shows the exact results, and the dashed line is that from AIA. In (b), the peak at $\omega/\Omega=2.5$ corresponds to the resonance $2\omega=\Delta_0$.  (c) Interferometric pattern using AIA: the long-time averaged population in the excited state ($\bar P_+$) as a function of $\Delta_0/\Omega$ and $\delta/\Omega$ for $\omega =0.32 \Omega$. The density peaks correspond to the resonances, and the solid lines mark the validity of AIA.}
	\label{fig:2}
\end{figure}

The basic idea of AIA is to divide the time evolution into adiabatic and non-adiabatic regimes as shown in Fig. \ref{fig:1}(b) \cite{ash07, she10, gar97}. In the adiabatic regime,  the system remains in the instantaneous eigenstate of the Hamiltonian, whereas in the non-adiabatic or impulse regime, the LZT takes place. In the LZ model, $\Delta(t)=vt$, where $v$ is the rate at which the detuning is varied across the avoided level-crossing \cite{lan32, zen32}. As seen in Fig. \ref{fig:1}(a), the energy gap between the two levels ($E_+$ and $E_-$) is maximum in the limit $t\to\pm \infty$ and is minimum at $t=0$ with a gap of $\Omega$. The system evolves adiabatically if $(E_+-E_-)^2/v\gg 1$ and  non-adiabatically otherwise \cite{vit99}. We approximately show the adiabatic and diabatic regimes in Fig. \ref{fig:1}(b) separated at the time $t'$, and while implementing AIA we take $t'\to 0$. Assuming  the atom is initially in the ground state, the transition probability to the excited state after a single-sweep across the avoided level crossing is,
\begin{equation}
P_{LZ} = \exp\left(-\pi\frac{\Omega^2}{2|v|}\right).
\label{plz}
\end{equation}
For a slow quench ($v\to 0$), the excitation probability is minimal ($P_{LZ}\to 0$), whereas, for a sudden ($v \to \infty$) one, there is a complete transition to the excited state ($P_{LZ}\to 1$). As shown in Fig. \ref{fig:1}(d), the exact dynamics are more involved, and the transition mostly takes place in the vicinity of the avoided level crossing, which constitutes the impulse region.   
 
Now, we consider the detuning periodic in time: $\Delta (t) = \Delta_0 +\delta \sin (\omega t)$, where $\delta$ and $\omega$ are the amplitude and the frequency of the modulation, respectively. In this case, the system is taken across the avoided level crossing ($\Delta(t)=0$) periodically at times $\tau_{2n}=[2n\pi+\sin^{-1}\left(-\Delta_0/\delta\right)]/\omega$ and $\tau_{2n+1}=[(2n+1)\pi-\sin^{-1}\left(-\Delta_0/\delta\right)]/\omega$ where $n=0, 1, 2, ...$. The adiabatic evolution between the two avoided level crossings is governed by the unitary matrix (written in the adiabatic basis $\{|\phi_+\rangle, |\phi_-\rangle\}$),
\begin{equation}
\hat{U}(t_2, t_1)= \left( {
	\begin{array}{ccc} 
	e^{-i\zeta_{+} }&  0 \nonumber \\
	0 & e^{-i\zeta_{-}}  
	\end{array} } \right)
\end{equation}
where $\zeta_\pm=\int_{t_1}^{t_2} dt E_\pm(t)$ are the accumulated dynamical phases. For non-zero bias ($\Delta_0\neq 0$), we have $\zeta_+\neq \zeta_-$, and the matrices $\hat{U}_1$ and $\hat{U}_2$, for the left and right sides of the crossings becomes non-identical [see Fig.~\ref{fig:1}(d)]. 

{\em Non-adiabatic evolution}. In the vicinity of the avoided level crossings, the detuning can be approximated as $\Delta(\tau_n\pm t) \approx \pm vt$ with $v  = \omega\sqrt{\delta^2 -\Delta_0^2}$ \cite{she10}. It makes the scenario identical to that of the LZ model, and we can use the result given in Eq.~(\ref{plz}). Eventually, we obtain the non-adiabatic LZT matrix in the adiabatic basis as,
\begin{equation}
\hat{G}_{LZ} 
= \left( {
	\begin{array}{ccc} 
	e^{-i\tilde{\phi}_s}\sqrt{1-P_{LZ}}  & -\sqrt{P_{LZ}} \\
	\sqrt{P_{LZ}} & e^{i\tilde{\phi}_s}\sqrt{1-P_{LZ}} 
	\end{array} } \right)
\label{LZM}
\end{equation}
where $\tilde{\phi}_s= \gamma(\ln \gamma -1)+ \arg \Gamma (1-i \gamma)+\frac{\pi}{4}$ is the Stokes phase with $\gamma=\Omega^2/4v$ being the adiabaticity parameter, and $\Gamma$ is the gamma function \cite{she10}. In terms of $\gamma$, the slow and sudden quenches are indicated respectively by $\gamma\gg 1$ and $\gamma\ll 1$. 

\subsection{Comparison with Multi-slit interference Pattern}
\label{ksl}
\begin{figure}
	\centering
	\includegraphics[width= .8\columnwidth]{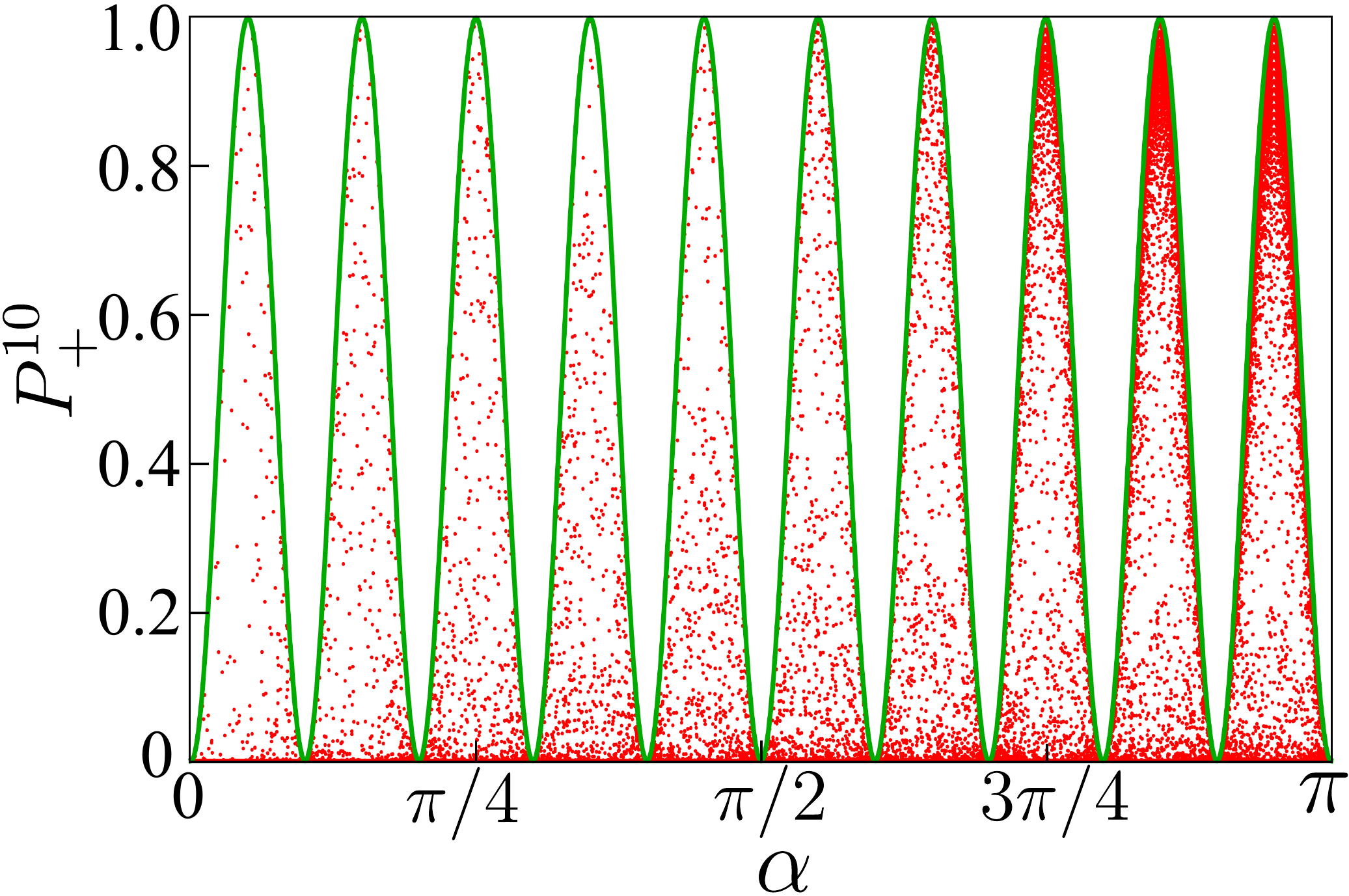}
	\caption{The excitation probability after ten cycles as a function of $\alpha$ for $\delta/\Omega=20$, $\Delta_0=0$ and $\omega$ is varied. A given $\alpha$ is not associated with a unique value of $\omega$, leading to the scattered points, but bounded by a maximum value of $\sin^2k\alpha$ shown by the solid line.}
	\label{fig:3}
\end{figure}

Over a half-cycle, say from $t_i=\tau_1-\pi/2\omega$ to $t_f=\tau_1+\pi/2\omega$, we can write the evolution matrix as $\hat U_2(t_f, \tau_1)\hat{G}^T_{LZ}\hat U_1(\tau_1, t_i)$. In general, the order of the transition and adiabatic matrices should be carefully chosen depending on  $\Delta_0$, the initial ($t_i$) and final $(t_f)$ times. Similarly, the evolution matrix for one complete cycle, and $t_i=0$ to $t_f=2\pi/\omega$ with $\Delta_0>0$ [see Figs. \ref{fig:1}(c) and \ref{fig:1}(d)] is $\hat F=\hat U_1(2\pi/\omega, \tau_2)\hat{G}_{LZ}\hat U_2(\tau_2, \tau_1)\hat{G}_{LZ}^T\hat U_1(\tau_1, 0)$, where the label $T$ stands for the transpose of the matrix. For the full cycle, the LZTs take-place at two instants. Writing the matrix as:
\begin{eqnarray}
\hat{F} = e^{i\phi_G}
  \left( {
	\begin{array}{ccc} 
	g_{11}&  -g_{21}^*  \\
	g_{21} & g_{11}^*
	\end{array} } \right)
\end{eqnarray}
where $\phi_G = \exp\left(i\int_{0}^{2\pi/\omega} \Delta(t) dt/2\right)$,
\begin{eqnarray}
g_{11} = e^{-i\eta_0}(1-P_{LZ}) +e^{-i\eta_1} P_{LZ} \\
g_{21} =(e^{-i\eta_3} - e^{-i\eta_2})e^{i\tilde{\phi}_s} \sqrt{(1-P_{LZ})P_{LZ}}
\end{eqnarray}
with $\eta_0 = \frac{1}{2}\int_{0}^{2\pi/\omega} \bar{\Omega} dt + 2\tilde{\phi}_s$, $\eta_1 = \frac{1}{2}\int_{0}^{2\pi/\omega} \bar{\Omega} dt-\int_{\tau_1}^{\tau_2} \bar{\Omega} dt$, $\eta_2 = \frac{1}{2}\int_{0}^{2\pi/\omega} \bar{\Omega} dt-\int_{\tau_2}^{2\pi/\omega} \bar{\Omega} dt+2\tilde{\phi}_s $ and  $\eta_3 =\int_{0}^{\tau_1} \bar{\Omega} dt- \frac{1}{2}\int_{0}^{2\pi/\omega} \bar{\Omega} dt$ being the dynamical phases.  Assuming the system is initially in the ground state, the transition probability to the excited state after one full cycle is, 
\begin{equation}
P_+^1=|g_{21}|^2=4(1-P_{LZ})P_{LZ}\sin^2\phi_s.
\label{tp1}
\end{equation}
Eq.~(\ref{tp1}) implies that the transition probability after one period is the result of the quantum interference between the transition amplitudes at $\tau_1$ and $\tau_2$, and also a periodic function of the phase $\phi_s=\frac{1}{2}\int_{\tau_1}^{\tau_2} \bar{\Omega} dt+\tilde{\phi}_s$, called the St\"uckelberg phase. Thus, the dynamical phase acquired between the LZTs at $\tau_1$ and $\tau_2$, and the phase change during the LZTs ($\tilde{\phi}_s$) become highly relevant to characterize the full cycle dynamics. We have constructive (destructive) interference with $|g_{21}|^2=P_{LZ}$ ($|g_{21}|^2=0$) when $\phi_s=(n+1/2)\pi$ ($\phi_s=n\pi$) where $n=0, 1, 2, ...$. As long as the LZT time (the duration for which the LZT takes place across an avoided crossing) is sufficiently shorter than the duration of adiabatic evolution between the two transitions, i.e., when $\tau_{LZ}<[\pi-2\sin^{-1}(-\Delta_0/\delta)]/\omega$ AIA is valid. The upper limit for $\tau_{LZ}$ is given by $\left(\sqrt{\gamma}/\Omega\right) \max(1, \gamma)$, and the validity of AIA requires $\delta-\Delta_0>\Omega$ and $\delta\omega>\Omega^2$ \cite{gar97, ash07, she10}. In Fig.~\ref{fig:2}(a), we show the transition probability to the excited state after a single cycle for $\delta = 20 \Omega$,  $t_i = \pi/2\omega$, $t_f = 5\pi/2\omega$ and $\Delta_0=5\Omega$ when the atom is initially prepared in the ground state. The results from AIA are found to be in an excellent agreement with the exact numerical results.

It is straightforward to extend AIA for multiple cycles, and we have $\hat F^k=(\hat{U}_1\hat{G}_{LZ}\hat U_2\hat{G}_{LZ}^T\hat{U}_1)^k$ for $k$-cycles. Writing it in the matrix form  \cite{she10}, 
\begin{eqnarray}
\hat{F}^k = e^{ik\phi_G}
  \left( {
	\begin{array}{ccc} 
	u_{11}&  -u_{21}^*  \\
	u_{21} & u_{11}^*
	\end{array} } \right),
\end{eqnarray}
where $u_{11}=\cos k\alpha+i\rm{Im}(g_{11})\sin k\alpha/\sin\alpha$ and $u_{21}=g_{21}\sin k\alpha/\sin\alpha$ with $\cos \alpha=\rm{Re}(g_{11})$. Therefore the transition probability from the ground to the excited state after $k$-cycles is 
\begin{equation}
P_+^k=|u_{21}|^2=4(1-P_{LZ})P_{LZ}\sin^2\phi_s\frac{\sin^2 k\alpha}{\sin^2 \alpha}
\label{nc2la}
\end{equation}
The long-time ($k\gg 1$) averaged occupation probability in the excited state is 
\begin{equation}
\bar{P}_+ = \frac{2(1-P_{LZ})P_{LZ}\sin^2\phi_s}{\sqrt{[4(1-P_{LZ})P_{LZ}\sin^2\phi_s]^2 + \text{Im}(g_{11})^2 }}.
\end{equation}
Thus, a complete resonant transition between the adiabatic states $( \bar{P}_+ = \bar{P}_- = 1/2)$ occurs when  $\text{Im}(g_{11}) = -[(P_{LZ})\sin{\eta_1}+ (1-P_{LZ})\sin{\eta_0} ]=0 $.
In the fast passage limit ($\gamma \gg 1$), $P_{LZ} \approx 1$, the resonance condition reduces to $\Delta_0 = n\omega$. The peak at $\omega/\Omega=2.5$ in Fig. \ref{fig:2}(b) is attributed to the resonance at $2\omega=\Delta_0$. In the slow passage limit a simple relation for the resonances are not possible, but can be identified from the density peaks of $\bar P_+$  [see Fig. \ref{fig:2}(c)] for smaller values of $\delta/\Omega$ \cite{sch12}. The resonances $\Delta_0 = n\omega$, also imply a coherent Rabi oscillations between the states, $|g\rangle$ and $|r\rangle$ \cite{ash07}.

\begin{figure}
	\centering
	\includegraphics[width= 1.\columnwidth]{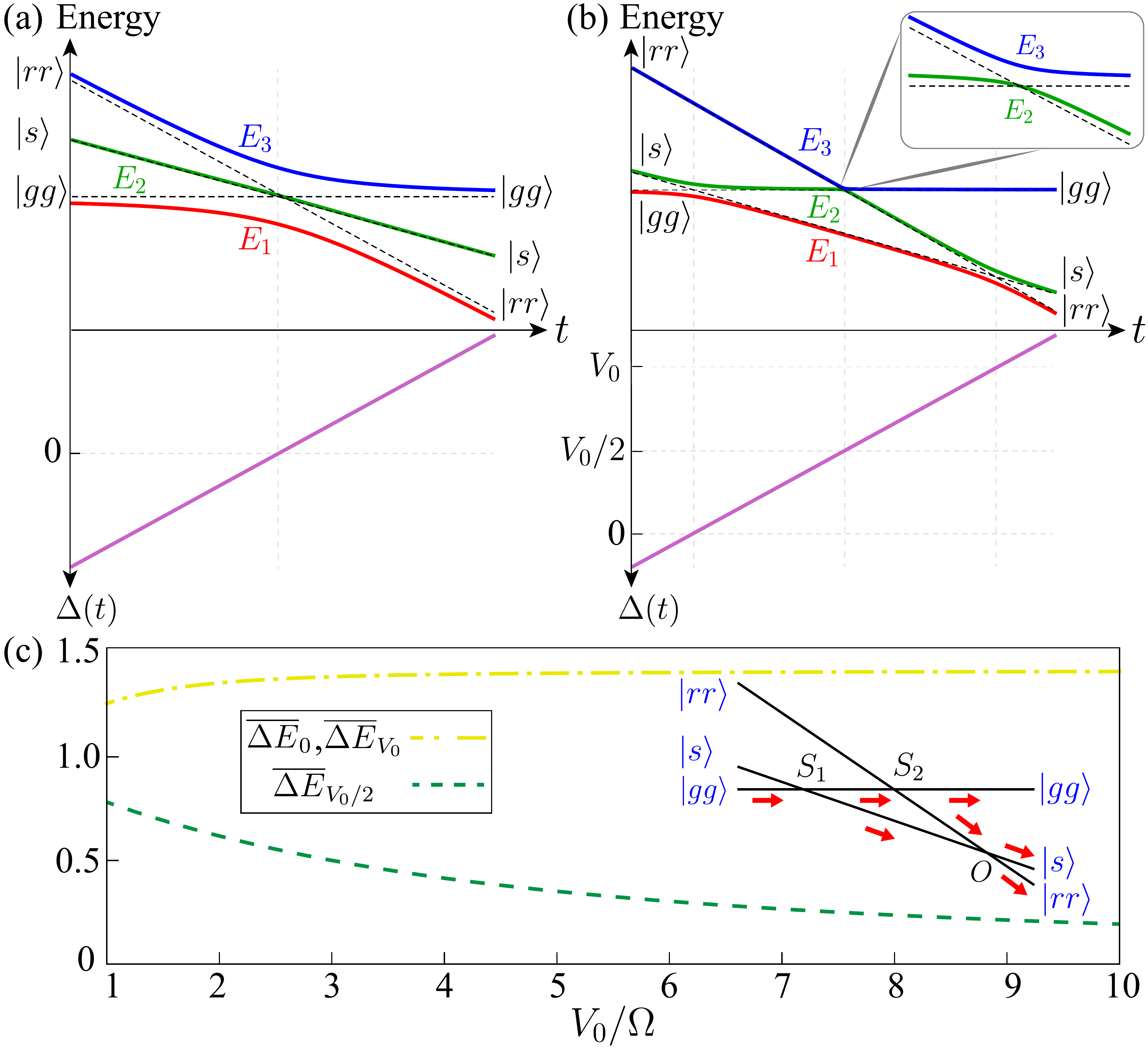}
	\caption{Energy eigenvalues $E_j$ as a function of the instantaneous detuning (the time dependence, $\Delta(t)=vt$ is shown at the bottom) for (a) $V_0=0.1\Omega$ and (b) $V_0=5\Omega$. The dashed lines show the diabatic energy levels. The avoided crossings in (b) form a triangular LZ model. The inset in (b) shows the avoided crossing at $V_0/2$. The asymptotic states at $t\to\pm\infty$ are given in the left and right end of the level diagrams. (c) shows the energy gaps $\overline {\Delta E}_{\alpha}=\Delta E_{\alpha}/\Omega$ at the avoided crossings as a function of $V_0$. The inset shows the schematic setup for the LZ interferometer in which the first ($S_1$) and the second ($S_2$) crossings act as beam splitters. At the last crossing $O$ mixing takes place. }
	\label{fig:4}
\end{figure}

We identify an interesting similarity in the form of the Eq.~(\ref{nc2la}) with the intensity distribution of an array of $k$ narrow equal amplitude slits (or an antenna array) interference pattern. For the latter case, the intensity along the direction $\theta$ is given by \cite{roy75}, 
\begin{equation}
I(\theta)=I_0\frac{\sin^2(k\phi/2)}{\sin^2(\phi/2)},
\label{slit}
\end{equation}
where $I_0$ is the intensity from a single slit. The angle, $\phi=2\pi d\sin\theta/\lambda$ is the phase difference between the consecutive slits where $d$ is the spacing between the adjacent slits, and $\lambda$ is the wavelength of light. Neglecting the slit widths ($I_0$ becomes a constant), the intensity pattern has principal maxima at  $\phi=2n\pi$ where $n=0, 1, 2, 3, ...$, and between two principal maxima there are $k-1$ minima located at $\phi/2=\pi/N, 2\pi/N, ..., (N-1)\pi/N$. Also, there  are $N-2$ secondary maxima between two principal maxima. Though the form of equations is the same, they exhibit significant differences. For instance, $I_0$ in Eq.~(\ref{slit}) does not depend on $\phi$, whereas the corresponding term $|g_{21}|^2$ in Eq.~(\ref{nc2la}) and $\alpha$ are not independent. In the latter case, a little algebra reveals that the maxima in the transition probability occur at $\cos k\alpha=0$ or $\alpha=(2n+1)\pi/2k$, and the minima occur at $\sin k\alpha=0$ or $\alpha=n\pi/k$. Thus, $\alpha=0$ doesn't correspond to a maximum but a minimum, in contrast with the antenna array intensity distribution for which $\phi=0$ represents a principal maximum. The other difference is that there are no secondary maxima in the excitation probability and consequently, one minimum between the maxima. It can be seen in Fig.~\ref{fig:3}, which shows the results from AIA for the excitation probability after ten cycles ($k=10$) as a function of $\alpha$ for $\delta/\Omega=20$, $\Delta_0=0$ and $\omega$ is varied (similar results can be obtained if $\delta$ or $\Delta_0$ is varied). There is no one to one correspondence between $\alpha$ and $\omega$, leading to scattered red dots in Fig.~\ref{fig:3}. For a fixed $\alpha$, the maximum value of $P_+^k$ is provided by the condition $\Im(g_{11})=0$, and we have $(P_+^k)_{MAX}=\sin^2k\alpha$, which is shown by the solid line in Fig.~\ref{fig:3}. As the number of cycles ($k$) increases, the number of peaks increases and also they get sharper. These results imply that, by correctly choosing the driving parameters and the number of cycles $k$, we can control the transition probability in a two-level atom or a qubit. The same results also hold for the periodically driven two-atom case, which will be discussed in Sec.~\ref{pd1}.


\section{Two Two-level Atoms: Rydberg-Rydberg interactions}
\label{2as}
 The two-atom setup has been a common scenario in several experimental studies \cite{beg13, gae09, wil10, ise10, rya10,rav14, lab14,rav15, jau15,del17,zen17,pic18,lev18} and in this case, the RRIs become relevant. The system is described by the Hamiltonian,
\begin{equation}
\hat H=-\Delta(t)\sum_{i=1}^2\hat\sigma_{rr}^{i}+\frac{\Omega}{2}\sum_{i=1}^2\hat\sigma_x^{i}+V_0\hat\sigma_{rr}^{1}\hat\sigma_{rr}^{2},
\label{h2}
\end{equation}
where $V_0 = C_6/R^6$ is the RRI between the atoms separated by a distance $R$ with $C_6$ being the van der Waals coefficient \cite{rei07}. For $V_0=0$, the two atoms are decoupled, and each of them exhibits independent LZ dynamics. To analyze the interacting case, we use the diabatic basis  $\{ \ket{gg}, \ket{s}, \ket{rr} \}$ where $\ket{s}=(\ket{gr}+\ket{rg})/\sqrt{2}$ is the symmetric state and the asymmetric state  $(\ket{gr}-\ket{rg})/\sqrt{2}$ can be disregarded in our study. The instantaneous eigenstates of $\hat H$ in the diabatic basis are
\begin{equation}
|j\rangle = \frac{1}{A}
\left( {
\begin{array}{ccc} 
-\frac{V_0-2\Delta(t)-E_{j}}{E_{j}} \\
-\frac{\sqrt{2}\left(V_0-2\Delta(t)-E_{j}\right)}{\Omega}  \\
 1
\end{array} } \right)
\end{equation}
where $j\in \{1,2,3\}$, $A$ is the normalization constant and the states $|j\rangle$ form the adiabatic basis. Thus, the two-atom setup effectively acts as a three-level system. Asymptotically the state $|j\rangle$ approaches the diabatic ones as $\lim_{\Delta\to-\infty} |1\rangle=|gg\rangle$, $\lim_{\Delta\to\infty} |1\rangle=|rr\rangle$, $\lim_{\Delta\to\pm\infty} |2\rangle=|s\rangle$, $\lim_{\Delta\to-\infty} |3\rangle=|rr\rangle$ and $\lim_{\Delta\to\infty} |3\rangle=|gg\rangle$. Upon diagonalizing the Hamiltonian, the instantaneous eigenenergies  $E_{j}$ are obtained as the roots of the cubic polynomial: $f(x) = -x^3 +(V_0-3\Delta)x^2+(V_0\Delta-2\Delta^2+\Omega^2)x -V_0\Omega^2 /2 +\Delta\Omega^2$ and we get
\begin{align}
E_n =  \frac{1 }{3}\left[ V_0-3\Delta +2|C| \cos(\theta_n/3)\right]
\end{align}
where $\theta_n = 3\arccos(\Re(C)/|C|)+ \lambda_n$ with $\lambda_n = 2(3-n)\pi$, $C= \left[\left(D_1 - \sqrt{D_1^2-4D_0^3}\right)/2\right]^{1/3}$, $D_0 = V_0^2-3V_0\Delta(t)+3\Delta(t)^2+3\Omega^2$, and $D_1= 2V_0^3 -9V_0^2\Delta(t)+9V_0\Delta(t)^2 -9V_0\Omega^2/2$. For sufficiently large $V_0$, the spectrum exhibits three distinct avoided level crossings, located at (i) $\Delta=0$ $\left(|1\rangle\leftrightarrow|2\rangle\right)$, (ii) $\Delta=V_0/2$ $\left(|2\rangle\leftrightarrow|3\rangle\right)$ and (iii) $\Delta=V_0$ $\left(|1\rangle\leftrightarrow|2\rangle\right)$, as seen in Fig.~\ref{fig:4}(b). 

The energy gaps $\Delta E_{\alpha\in\{0, V_0/2, V_0\}}$ at the avoided crossings, [as a function of $V_0$ are shown in Fig.~\ref{fig:4}(c)] are very relevant in LZ dynamics. We have $\Delta E_0=\Delta E_{V_0}$, which increases with $V_0$ and eventually saturates to $\sqrt{2}\Omega$ at large $V_0$. Whereas $\Delta E_{V_0/2}$ decreases inversely with $V_0$, i.e., $\Delta E_{V_0/2}\sim 1/V_0$. The vanishingly small $\Delta E_{V_0/2}$ at large $V_0$ can be associated with the fact that $|gg\rangle$ and $|rr\rangle$ are not directly coupled. Note that, a sufficiently large $V_0$ can isolate the different avoided crossings from each other.
\begin{figure*}
	\centering
	\includegraphics[width= 2\columnwidth]{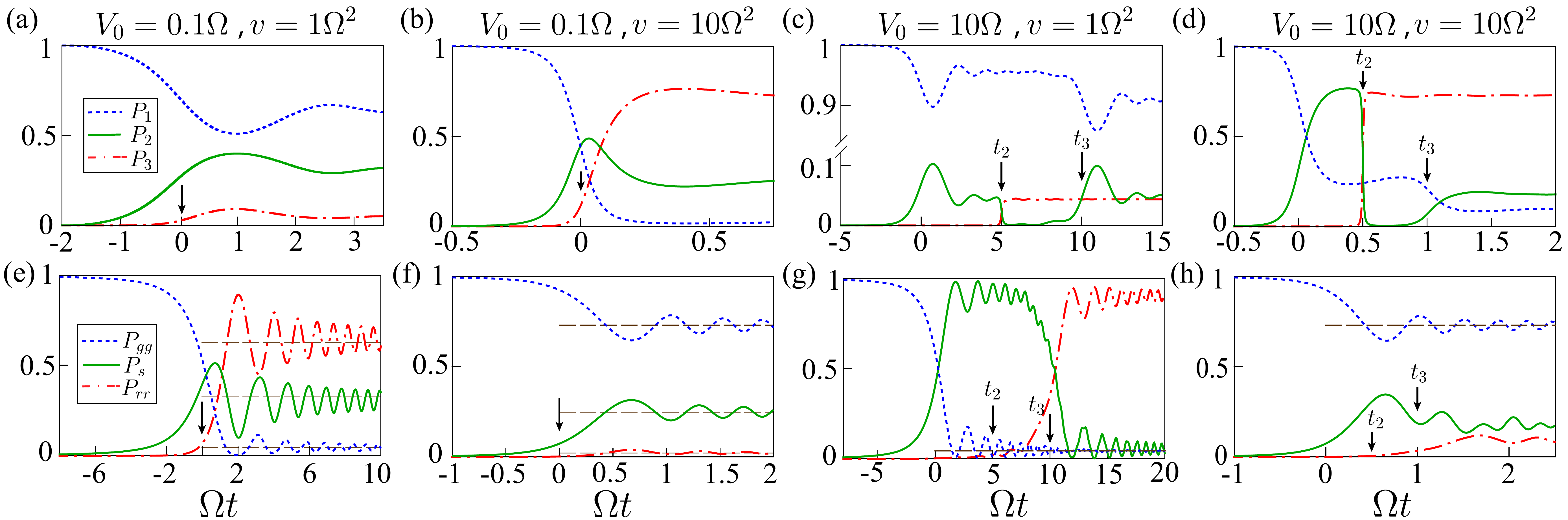}
	\caption{The dynamics of populations in the adiabatic (a)-(d) and the diabatic (e)-(h) states for the initial state $|\psi(t_i)\rangle=|1\rangle\sim |gg\rangle$, $v= (1\Omega^2, 10\Omega^2$), and $V_0= (0.1\Omega, 10\Omega)$. The first LZT takes-place in the vicinity of $t=0$. The thin arrows show the times around which the second ($t_2$) and the third ($t_3$) LZTs occur. In (a), (b), (e), and (f), the LZTs are not resolvable, so a single arrow is shown. The dashed horizontal lines in the bottom row show the results from the non-interacting model.}
	\label{fig:5}
\end{figure*}

{\em Different LZ Models}.--- Among the diabatic states, $|s\rangle$ couples to both $|gg\rangle$ and $|rr\rangle$, but $|gg\rangle$ and $|rr\rangle$ are not coupled to each other. Therefore, for vanishing interactions the two atom setup converges to a three level bow-tie LZ model \cite{car86, ost97, dem00, dem01}. The same two atom setup can mimic a  four level bow-tie model if an offset in Rabi frequencies or detunings is provided between two atoms \cite{sri19, mil19}. For sufficiently large $V_0$ (blockade regime), the avoided level crossings form a triangular geometry [see Fig.~\ref{fig:4}(b)]. A triangle LZ model is known to exhibit beats and step patterns in the population dynamics \cite{kis13, par18}. The Hamiltonian in Eq.~(\ref{h2}) can be written as an SU(3) model using the mapping: $\{ \ket{gg}, \ket{s}, \ket{rr} \} \to \{ \ket{+1}, \ket{0}, \ket{-1} \}$, i.e., \cite{ken16, kis13} 
\begin{equation}
\hat H_s=\left[\Delta(t)-\frac{V_0}{2}\right]\hat S_z+\Omega\hat S_x+ \frac{V_0}{2}\hat S_z^2,
\label{sh2}
\end{equation}
where $\hat S_z$ and $\hat S_x$ are the spin-1 matrices, and the last term is known as the easy-axis single-ion anisotropy in the context of magnetic systems. In the limit $V_0\to 0$, the three avoided level crossings merge at the point of zero detuning [see Fig. \ref{fig:4}(a)], and we get a spin-1 SU(2) model \cite{ban19}. The presence of RRI makes the model in Eq.~(\ref{sh2}) non-linear in SU(2) basis, but the nonlinearity can be removed by expressing in terms of the generators (Gell-Mann matrices) of the SU(3) group \cite{kis13}. 


\subsection{Three-level Landau-Zener Model}
\label{lzm}

In the three-level LZ model, the detuning varies linearly in time \cite{car86,shy04,ban19} and the Hamiltonian in the diabatic basis $\{ \ket{gg}, \ket{s}, \ket{rr} \}$ is given by,
\begin{equation}
\hat{H} 
= \left( {
	\begin{array}{ccc} 
	0 & \frac{\Omega}{\sqrt{2}} & 0  \\
	\frac{\Omega}{\sqrt{2}} & -vt & \frac{\Omega}{\sqrt{2}}  \\
	0 & \frac{\Omega}{\sqrt{2}} &  -2vt+V_0
	\end{array} } \right).
\label{H3}
\end{equation}
We consider a linear sweep from far left to far right including all the three avoided level crossings, and analyze the LZ dynamics as a function of both $v$ and $V_0$ for different initial states. We set the initial ($t_i$) and final ($t_f$) time such that the adiabatic states converge to the diabatic ones. The first LZT takes-place from $|1\rangle$ to $|2\rangle$ at around the time $t_1=0$, the second one from $|2\rangle$ to $|3\rangle$ around $t_2=V_0/2v$ and the last one is between $|1\rangle$ and $|2\rangle$ around $t_3=V_0/v$. The state $|3\rangle$ is involved only in one LZT (the second one), whereas $|1\rangle$ and $|2\rangle$ are part of more than one LZTs. The latter implies that the final population in $|1\rangle$ and $|2\rangle$, i.e., $P_1(t_f)$ and $P_2(t_f)$, is determined by the interference of distinct LZTs. Also, using simple scaling arguments (defining $\tilde t=t/\sqrt{v}$ in the Schr\"odinger equation), we can argue that the transition probabilities will only be a function of two parameters: $\Omega/\sqrt{v}$ and $V_0/\sqrt{v}$. Below, we discuss the dynamics for three different initial states.

\subsubsection{$Initial state: |\psi(t_i)\rangle=|1\rangle$}
\begin{figure}
	\centering
	\includegraphics[width=1.\columnwidth]{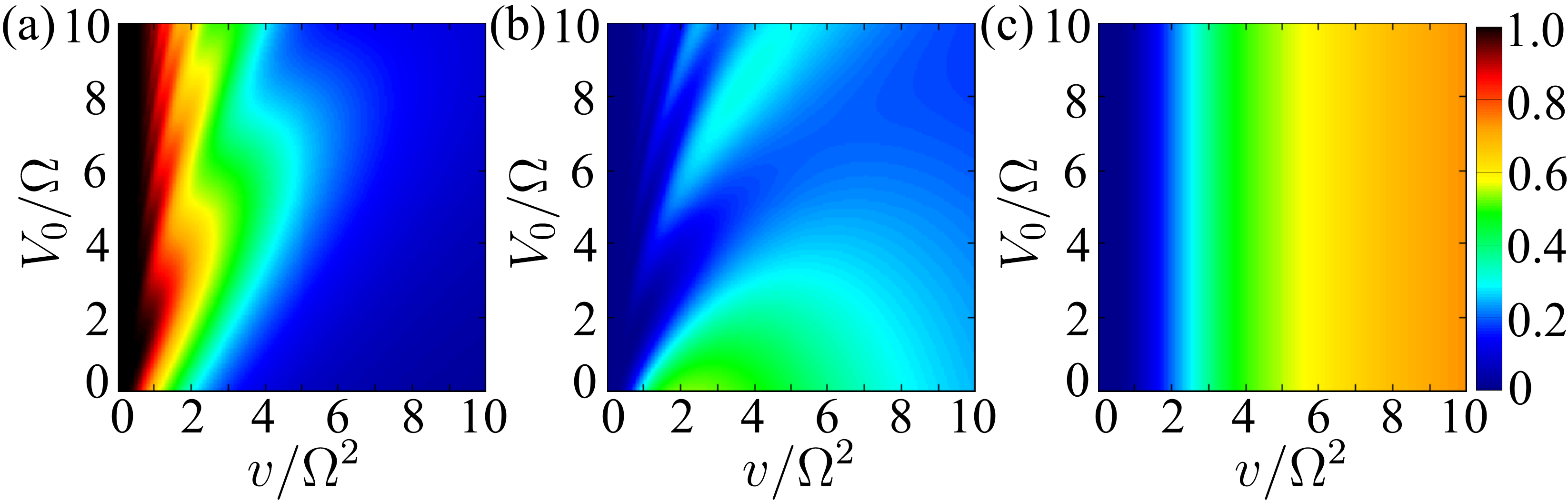}
	\caption{The final population in (a) $|1\rangle\sim|rr\rangle$, (b) $|2\rangle\sim|s\rangle$ and (c) $|3\rangle\sim|gg\rangle$, after the linear quench, as a function of $v$ and $V_0$ for the initial state $|\psi(t_i)\rangle=|1\rangle$, $\Delta(t_i)=-10\Omega$, and $\Delta(t_f)=30\Omega+10v/\Omega$.}
	\label{fig:6}
\end{figure}

{\em Adiabatic states}. The  population dynamics in the adiabatic states is shown in Figs.~\ref{fig:5}(a)-\ref{fig:5}(d) for the initial state, $|\psi(t_i)\rangle=|1\rangle$, and $t_i$ is such that $|1\rangle\sim |gg\rangle$. For $V_0\ll\Omega$, the avoided level crossings are closely spaced [see Fig.~\ref{fig:4}(a)], and hence, both $|2\rangle$ and $|3\rangle$ get populated almost simultaneously if $v$ is sufficiently large, as seen in Figs.~\ref{fig:5}(a)-\ref{fig:5}(b). For $V_0\gg\Omega$, we can resolve the three LZTs in the dynamics as long as $V_0/\sqrt{v}$ is sufficiently large [see Figs. \ref{fig:5}(c)-\ref{fig:5}(d)]. The first and third transitions are seen as two major dips in $P_1(t)$, and they correspond to the population transfer from $|1\rangle$ to $|2\rangle$, which takes place at around $t\sim 0$ and $t_3$, respectively. Once the LZTs are resolved, we have a basic setup for the LZ interferometer based on amplitude splitting. It is schematically shown in the inset of Fig.~\ref{fig:4}(c). The (avoided) crossings play the role of beam splitters [$S_1$ and $S_2$ in Fig.~\ref{fig:4}(c)], and the energy gaps and the rate $v$ can be related to their thickness. At the last crossing $O$, the mixing takes place, and the final population in $|gg\rangle$ is taken to be the leakage from the interferometer.

Fig.~\ref{fig:6} shows the final population in the adiabatic/diabatic states. As expected, for a fixed $V_0$, larger $v$ leads to higher transition probabilities, which results in a smaller $P_1(t_f)$ and a larger $P_3(t_f)$. But, $P_2(t_f)$ displays a non-monotonous behavior. A better understanding of $P_j(t_f)$ can be obtained by accessing the dynamics across each avoided crossings separately. For instance, after the first avoided crossing, $P_2(t)$ becomes independent of $V_0$ for large $V_0$, but depends on $v$. This is because $\Delta E_0$ saturates to $\sqrt{2}\Omega$ at large $V_0$ [see Fig. \ref{fig:4}(c)]. At the same time, $\Delta E_{V_0/2}$ decreases and becomes significantly small ($\Delta E_{V_0/2}\ll\Omega$) at large $V_0$. The latter results in a complete and a sharp transition from $|2\rangle$ to $|3\rangle$ at the second LZT for sufficiently large $v$ [see the dashed-dotted line in Fig.~\ref{fig:5}(d)]. As a result, $P_3(t_f)$ becomes independent of $V_0$ at sufficiently large $V_0$ and only depends on $v$ [see Fig. \ref{fig:6}(c)]. Counter-intuitively, even for small $V_0$, we see that $P_3(t_f)$ is independent of $V_0$, which is better explained using the dynamics in the diabatic states (see below). 

{\em Diabatic states}. In Fig.~\ref{fig:5}(e)-\ref{fig:5}(h), we show the population dynamics in the diabatic states for the same  in Fig.~\ref{fig:5}(a)-\ref{fig:5}(d). In contrary to the adiabatic states, the population in the diabatic states exhibit clean oscillations (akin to Rabi oscillations) with the amplitude being damped over time [see Figs.~\ref{fig:5}(e)-\ref{fig:5}(h)] \cite{vit99}. The frequency of these oscillations increases over time since the effective instantaneous Rabi frequency increases with an increase in the detuning. For small values of $V_0$ and $v$, the amplitude of oscillation is larger. The reasons are two-fold, first, for small $V_0$, the three LZTs are closely placed, and second, having a small $v$, the system spends more time in the impulse regime. In the adiabatic limit ($v\ll\Omega^2$), after the sweep, the initial population in $|gg\rangle$ gets completely transfer to $|rr\rangle$ independent of the value of $V_0$ [see Fig.~\ref{fig:6}]. As $v$ increases, there is non-zero population in both $|s\rangle$ and $|rr\rangle$ states. With further increase in $v$, the transition between the diabatic states gets suppressed, reducing the final population in both $|s\rangle$ and $|rr\rangle$. 
Now, we consider weakly and strongly interacting cases separately.

\begin{figure}
	\centering
	\includegraphics[width=1.\columnwidth]{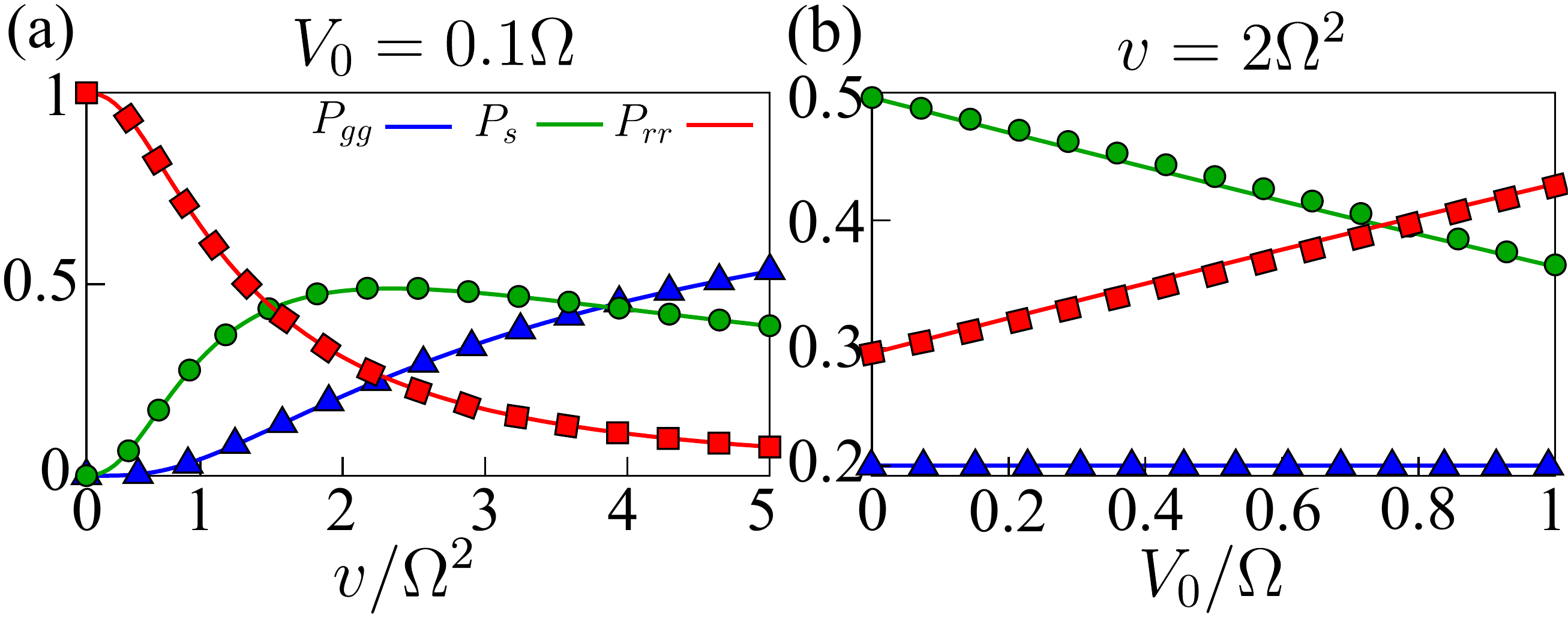}
	\caption{(a) The final population in the adiabatic/diabatic states as a function of $v$ for $V_0=0.1 \Omega$ with the initial state $|1\rangle\sim |gg\rangle$. (b) The same as in (a) but as a function of $V_0$ for $v=2\Omega^2$. The solid lines show exact results, and the filled squares, circles, and triangles are the theoretical prediction for small $V_0$ in the limit $t_f\to \infty$.} 
	\label{fig:7}
\end{figure}

{\em Weakly interacting case}.--- For $V_0\ll\Omega$ and $V_0/\sqrt{v}\ll 1$, the population get transfer to both $|s\rangle$ and $|rr\rangle$ at around $t\sim 0$, and the system does not spend significant time across the avoided crossings making the effect of interactions minimal. In this case, we can assume the atoms to be non-interacting, and we have $P_3(t\to \infty)=P_{gg}\sim P_{LZ}^2$, $P_2(t\to \infty)=P_{s}\sim 2P_{LZ}(1-P_{LZ})$ and $P_1(t\to \infty)=P_{rr}\sim (1-P_{LZ})^2$ where $P_{LZ}$ is given in Eq. (\ref{plz}). Dashed horizontal lines in Figs.~\ref{fig:5}(e)-\ref{fig:5}(h) show the results from the non-interacting approximation and are in good agreement with the numerical results. As $v$ gets smaller, $V_0$ introduces small corrections to the non-interacting results. From the numerical results, we see that $P_{gg}(t_f)$ is independent of $V_0$ [see fig.~\ref{fig:6}(c)], and therefore, we simply have $P_{gg}(t\to\infty)=P_{LZ}^2$. Based on the scaling arguments and insights from the numerical results, we can write down
\begin{equation}
P_s(t\to \infty)\sim 1-P_{LZ}^2 - (1-Q_{LZ})^2,
\label{ps}
\end{equation}
and $P_{rr}(t\to \infty)\sim (1-Q_{LZ}^2)$ where 
\begin{equation}
Q_{LZ}=P_{LZ}\exp\left(-\frac{\pi\Omega^2V_0}{4v^{3/2}}\right).
\label{qlz}
\end{equation}
These results are in an excellent agreement with the exact results for $P_j(t_f)$ [see Fig.~\ref{fig:7}], even for sufficiently large values of $V_0$. 

\begin{figure*}
	\centering
	\includegraphics[width= 2.\columnwidth]{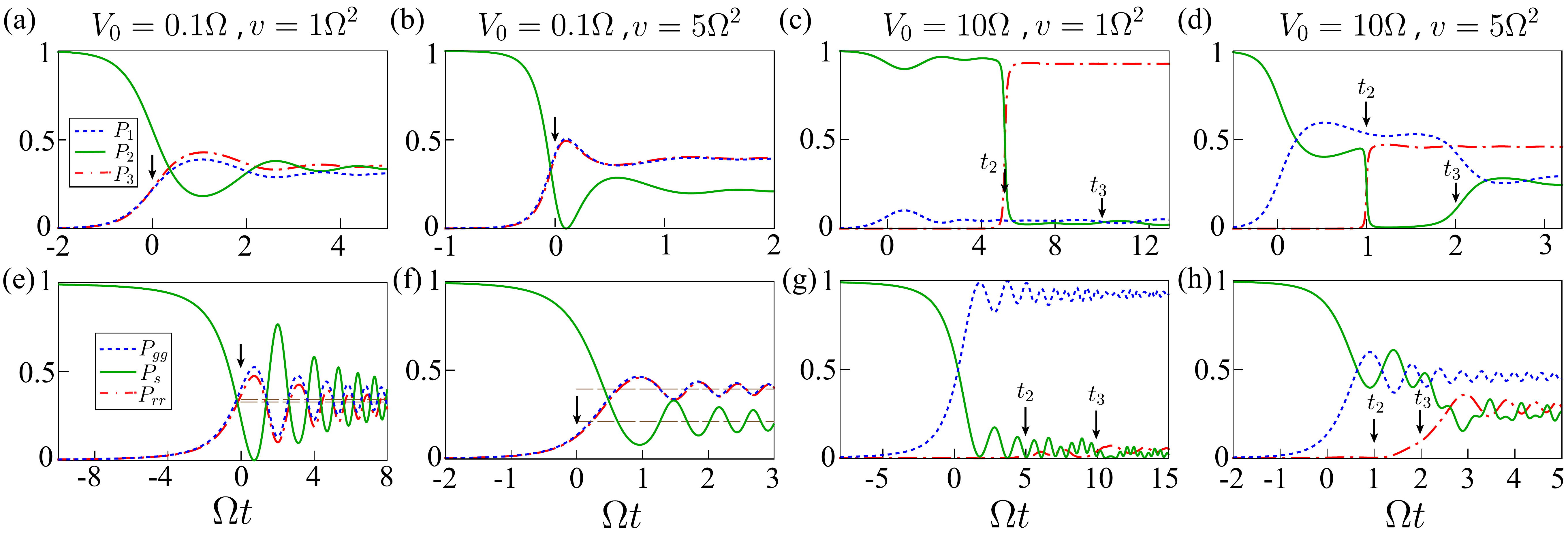}
	\caption{The dynamics of populations in the adiabatic (a)-(d) and the diabatic (e)-(h) states for the initial state $|\psi(t_i)\rangle=|2\rangle\sim |gg\rangle$, $v= (1\Omega^2, 5\Omega^2$), and $V_0= (0.1\Omega, 10\Omega)$.  The first LZT takes-place in the vicinity of $t=0$. The thin arrows show the times around which the second ($t_2$) and the third ($t_3$) LZTs occur. In (a), (b), (e), and (f), the LZTs are not resolvable, so a single arrow is shown. The dashed horizontal lines in (e) and (f) show the results from the non-interacting model.}
	\label{fig:8}
\end{figure*}

{\em Strongly interacting case}.--- For $V_0\gg\Omega$, the three avoided crossings are well separated, but the dynamics in the diabatic states do not show signatures of all three LZTs. The reason is that $|gg\rangle$ is not directly coupled to $|rr\rangle$ leaving no sign of second LZT (around $t_2$) in the dynamics [see Figs.~\ref{fig:5}(g) and \ref{fig:5}(h)]. Around the first LZT ($t\sim 0$), $P_{gg}$ decreases, and $P_s$ increases. If $v$ is sufficiently small, almost a complete transfer from $|gg\rangle$ to $|s\rangle$ takes place. As $\Delta(t)$ approaches the third LZT at around $t_3$, we have a population transfer from $|s\rangle$ to $|rr\rangle$. Thus, if $V_0/v\gg 1/\Omega$, the system evolves from an uncorrelated state ($|gg\rangle$), and transit through an entangled state ($|s\rangle$) and eventually settle in the (uncorrelated) doubly excited state ($|rr\rangle$). The duration in which the system stays in each of these states can be controlled via both $v$ and $V_0$. 

Previously, we have seen that $P_3(t_f)\sim P_{gg}$ is independent of $V_0$ [see Fig.~\ref{fig:6}(c)]. This feature has been explained for large $V_0$ using the adiabatic basis. A simple and complete picture, irrespective of $V_0$, can be obtained using diabatic states. The state $|gg\rangle$ is coupled only to $|s\rangle$, and hence, $V_0$ has no role in determining how much population transfer takes place from $|gg\rangle$ to $|s\rangle$. But, $V_0$ can affect the reverse since $|s\rangle$ is also coupled to $|rr\rangle$. In a single sweep and all the population initially in $|gg\rangle$, the reverse process is absent, leaving the final population in $|gg\rangle$ independent of $V_0$. Similarly, as we see later, if the initial state is $|rr\rangle$, the population  $P_1(t_f)\sim P_{rr}$ becomes independent of $V_0$.

\subsubsection{Initial state: $|\psi(t_i)\rangle=|2\rangle$}

 At this point, we comment briefly on the adiabaticity criteria.  When the initial state is $|1\rangle$ and for sufficiently large $V_0$, the gap $\Delta E_{0}\sim \sqrt{2}\Omega$ (same as $\Delta E_{V_0}$) sets the adiabatic limit, and is independent of $V_0$. Whereas, if the initial state is either $|2\rangle$ or $|3\rangle$, the adiabatic limit is determined by $\Delta E_{V_0/2}$ (smallest among the three gaps), which decreases monotonously with $V_0$ as seen in Fig. \ref{fig:4}(c). Therefore, for large $V_0$, when $\Delta(t)=V_0/2$, there is almost a complete population transfer between the states $|2\rangle$ and $|3\rangle$ unless $v$ is negligibly small. In other words, a large value of $V_0/\sqrt{v}$ may not guarantee an adiabatic evolution if the initial state is $|2\rangle$ or $|3\rangle$. For $V_0\gg\Omega$, approximating each avoided level crossings composed of only two levels and using the adiabatic theorem, we require $v\ll 2\Omega^2$ for an adiabatic evolution with the initial state $|1\rangle$. Similarly, we require $v\ll 4\Omega^4/V_0^2$ for an adiabatic evolution if the initial state is $|2\rangle$ or $|3\rangle$.

Fig.~\ref{fig:8} shows the population dynamics in both adiabatic and diabatic states for the initial state $|2\rangle$. For $V_0/\sqrt{v}\ll 1$, the interaction $V_0$ is irrelevant, and we have $P_1(t)=P_3(t)$ [see Fig.~\ref{fig:8}(b)]. Keeping $V_0\ll\Omega$, and for sufficiently small $v$, RRIs introduce an offset in the dynamics of the states $|1\rangle$ and  $|3\rangle$, i.e., $P_1(t)\neq P_3(t)$ [see Fig. \ref{fig:8}(a)]. For large $V_0$, the population from $|2\rangle$ first gets transferred to $|1\rangle$ at around $t=0$ [see Figs. \ref{fig:8}(c) and \ref{fig:8}(d)]. The remaining population in $|2\rangle$ gets completely transferred to $|3\rangle$ after the second avoided crossing. At around $t_3$ when the system crosses the third avoided crossing, the state $|2\rangle$ gains population from $|1\rangle$. Therefore, the final population in $|2\rangle$ increases with $v$ whereas that of $|1\rangle$ and $|3\rangle$ decreases.

Concerning the diabatic states, initially, the system is prepared in the $|s\rangle$ state. For $V_0\ll\Omega$, and sufficiently large $v$, the population in $|s\rangle$ gets transferred to $|gg\rangle$ and $|rr\rangle$ states symmetrically, as seen in Fig.~\ref{fig:8}(f). In this case, from the non-interacting LZ model, we have $P_{gg}(t\to\infty)=P_{rr}(t\to\infty)\sim 2P_{LZ} (1-P_{LZ} )$ and $P_s(t\to\infty)\sim 1- 4P_{LZ} (1-P_{LZ})$, which have been shown as horizontal lines in Figs.~\ref{fig:8}(e) and \ref{fig:8}(f) that are valid for $V_0/\sqrt{v}\ll 1$. Incorporating the effect of finite $V_0$ but still small, we get,

\begin{eqnarray}
P_{rr}(t\to\infty)\sim 1-P_{LZ}^2 - (1-R_{LZ})^2, \\
P_{gg}(t\to\infty)\sim 1-P_{LZ}^2 - (1-Q_{LZ})^2,
\end{eqnarray}
where $$R_{LZ}=P_{LZ}\exp\left(-\frac{\pi\Omega^2V_0}{2^{5/2}v^{3/2}}\right),$$ and $P_{s}(t\to\infty)=1-P_{rr}-P_{gg}$. These results are in excellent agreement with the exact results (not shown). 

For $V_0\gg\Omega$, the first population transfer takes place around $t=0$ to $|gg\rangle$ as shown in Figs.~\ref{fig:8}(g) and \ref{fig:8}(h). The second LZT is inactive since $|gg\rangle$ and $|rr\rangle$ are not directly coupled, leaving no sign in the dynamics. At around $t_3$, the population gets transfer from $|s\rangle$ to $|rr\rangle$. If the evolution across the first avoided crossing is entirely adiabatic, the system finally ends up in $|gg\rangle$, a state having no Rydberg excitations. This de-excitation is in stark contrast to dynamical creation of excitations by adiabatically sweeping the detuning from negative to large positive values \cite{poh10,van11,sch15}. 
 
\begin{figure}[hbt]
	\centering
	\includegraphics[width=1.\columnwidth]{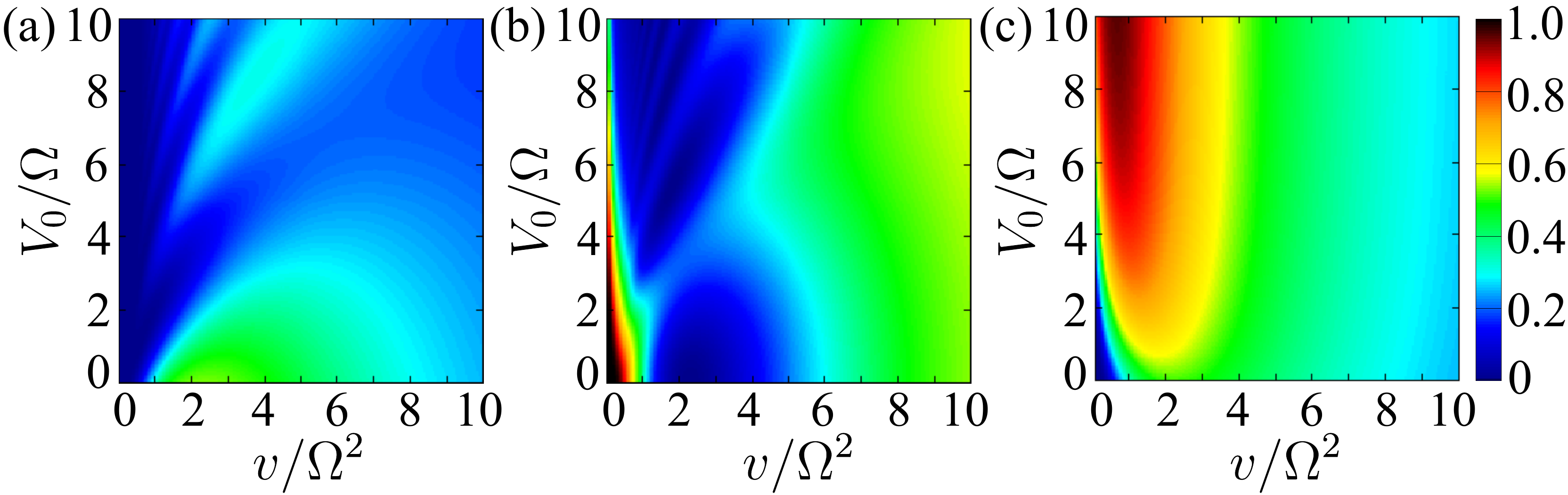}
	\caption{The final population in (a) $|1\rangle\sim|rr\rangle$, (b) $|2\rangle\sim|s\rangle$ and (c) $|3\rangle\sim|gg\rangle$, after the linear quench, as a function of $v$ and $V_0$ for the initial state $|\psi(t_i)\rangle=|2\rangle$, $\Delta(t_i)=-10\Omega$, and $\Delta(t_f)=30\Omega+10v/\Omega$.}
	\label{fig:9}
\end{figure}

\begin{figure}
	\centering
	\includegraphics[width= 1.\columnwidth]{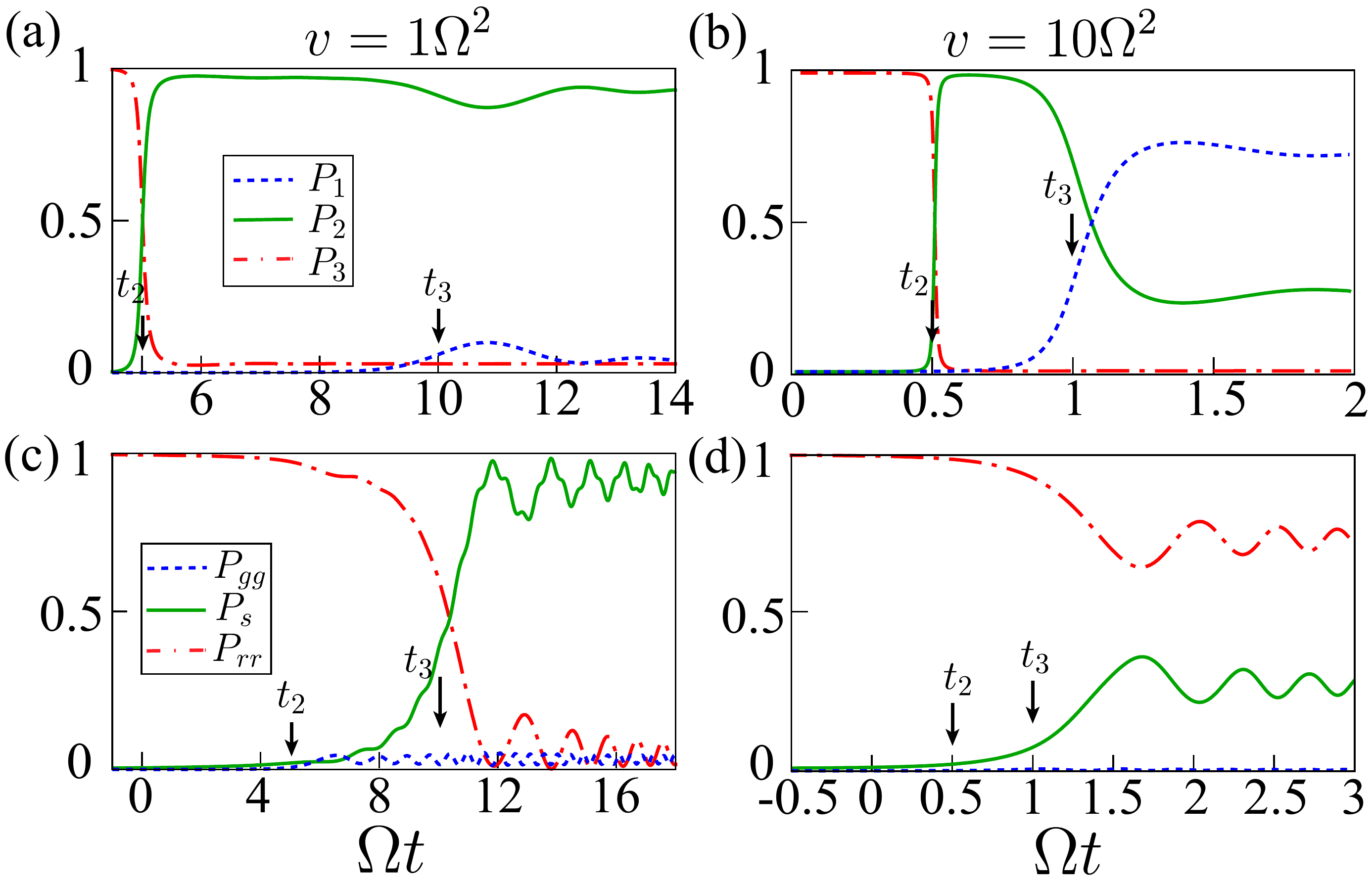}
	\caption{The dynamics of population in the adiabatic (a)-(b) and diabatic (c)-(d) states for the initial state $|\psi(t_i)\rangle=|3\rangle\sim |rr\rangle$ with different values of $v$ and $V_0=10\Omega$. The thin arrows show the times around which the second ($t_2$) and the third ($t_3$) LZTs occur.}
	\label{fig:10}
\end{figure}

The final population in the adiabatic/diabatic states as a function of $v$, and $V_0$ for the initial state $|2\rangle$ is shown in Fig.~\ref{fig:9}. In contrary to the case of initial state $|1\rangle$, here $P_3(t_f)\sim P_{gg}$ depends on $V_0$ [see Fig.~\ref{fig:9}(c)]. Another feature is that for large $V_0$, the population $P_3(t_f)$ depends non-monotonously on $v$. At small $v$, $P_3(t_f)$ increases with $v$, due to the smallness of $\Delta E_{V_0/2}$. Whereas at large values of $v$, across the first avoided crossing the transition amplitude increases with $v$ leading to a decrease in $P_{3}(t_f)$. The non-trivial patterns in $P_{1}(t_f)$ and $P_{2}(t_f)$ are due to the interference of LZTs at the different avoided crossings [see Figs.~\ref{fig:9}(a) and \ref{fig:9}(b)].

\subsubsection{$|\psi(t_i)\rangle=|3\rangle$}

\begin{figure}
	\centering
	\includegraphics[width=1.\columnwidth]{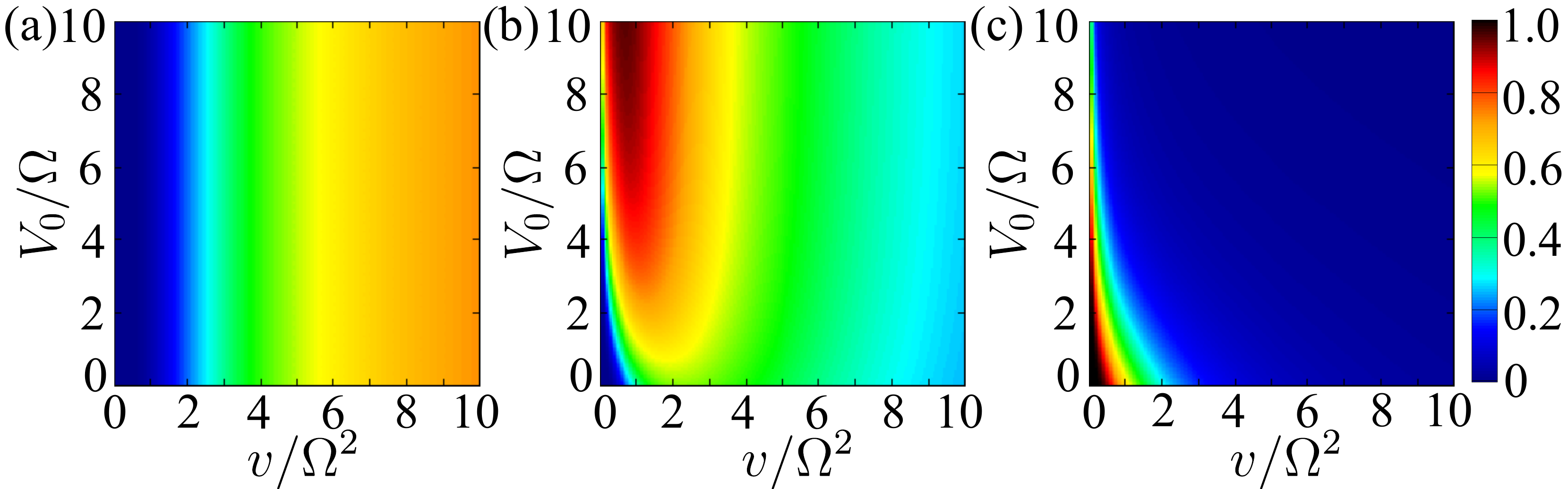}
	\caption{The final population in (a) $|1\rangle\sim|rr\rangle$, (b) $|2\rangle\sim|s\rangle$ and (c) $|3\rangle\sim|gg\rangle$, after the linear quench, as a function of $v$ and $V_0$ for the initial state $|\psi(t_i)\rangle=|2\rangle$, $\Delta(t_i)=-10\Omega$, and $\Delta(t_f)=30\Omega+10v/\Omega$.}
	\label{fig:11}
\end{figure}
For the initial state $|3\rangle$ and $V_0\gg\Omega$, the first avoided crossing is irrelevant in the dynamics. In this case, the transition first takes place at around $t_2$ to $|2\rangle$ [see Figs.~\ref{fig:10}(a) and \ref{fig:10}(b)]. Then, across the third avoided crossing, there is a transition from $|2\rangle$ to $|1\rangle$. Thus, for sufficiently large values of $V_0$ and $v$, we have $P_3(t_f)\sim 0$  [see Figs.~\ref{fig:10}(a) and \ref{fig:10}(b)]. Regarding the diabatic states, the initial population is solely in $|rr\rangle$. In this case, for $V_0\gg\Omega$, only the third avoided crossing is relevant, and the population can only transfer to $|s\rangle$ at around $t_3$ [see Figs.~\ref{fig:10}(c) and \ref{fig:10}(d)]. Also, larger the magnitude of $v$, the weaker the transition between $|rr\rangle$ and $|s\rangle$. For $V_0\ll\Omega$ and large $v$, we have the results for the final population using the non-interacting LZ model: $P_{rr}\sim P_{LZ}^2$, $P_{s}\sim 2P_{LZ}(1-P_{LZ})$, and $P_{gg}\sim (1-P_{LZ})^2$. After incorporating the effect of a finite $V_0$, we have $P_3(t_f\to\infty)=P_{gg}\sim (1-R_{LZ})^2$, $P_2(t_f\to\infty)=P_{s}\sim 1-P_{LZ}^2 - (1-R_{LZ})^2$, and $P_1(t_f\to\infty)=P_{rr}\sim P_{LZ}^2$.

The final population in the adiabatic/diabatic states as a function of $v$, and $V_0$ for the initial state $|3\rangle$ is shown in Fig.~\ref{fig:11}, and comparing it with Fig.~\ref{fig:6} and Fig.~\ref{fig:9}, we see that the patterns of final population in the $v-V_0$ plane are repeating. The identical patterns are (i) $P_3(t_f)$ with initial state $|1\rangle$ and $P_1(t_f)$ with initial state $|3\rangle$, (ii)$P_2(t_f)$ with initial state $|1\rangle$ and $P_1(t_f)$ with initial state $|2\rangle$ and (iii) $P_3(t_f)$ with initial state $|2\rangle$ and $P_2(t_f)$ with initial state $|3\rangle$. This implies that RRIs do not break the symmetry completely while swapping the states in the LZ model.

\subsubsection{Beats}
\label{bea}
 Depending on the geometric size of the triangle formed by the three avoided crossings [see Fig. \ref{fig:4}(b)], the triangular LZ model known to exhibit beats and step patterns in the population dynamics of the diabatic states \cite{kis13, fey15}. These patterns arise due to the quantum interference of distinct LZTs. We only briefly comment on the beats pattern in our setup.  The beats pattern is observed only in the population of the $P_s(t)$ as shown in Fig. \ref{fig:12}(a) for different initial states. Based on the calculations in Ref. \cite{kis13}, we would expect a beat pattern in $P_s(t)$ if $\Omega^2/4v\ll1$ and $V_0^2/4v<1$. The envelope frequency is found to be $V_0/2$, and the fast oscillation frequency changes over time as approximately $vt/4$.

\subsubsection{AIA}
\label{aia2}

\begin{figure}
	\centering
	\includegraphics[width=.8\columnwidth]{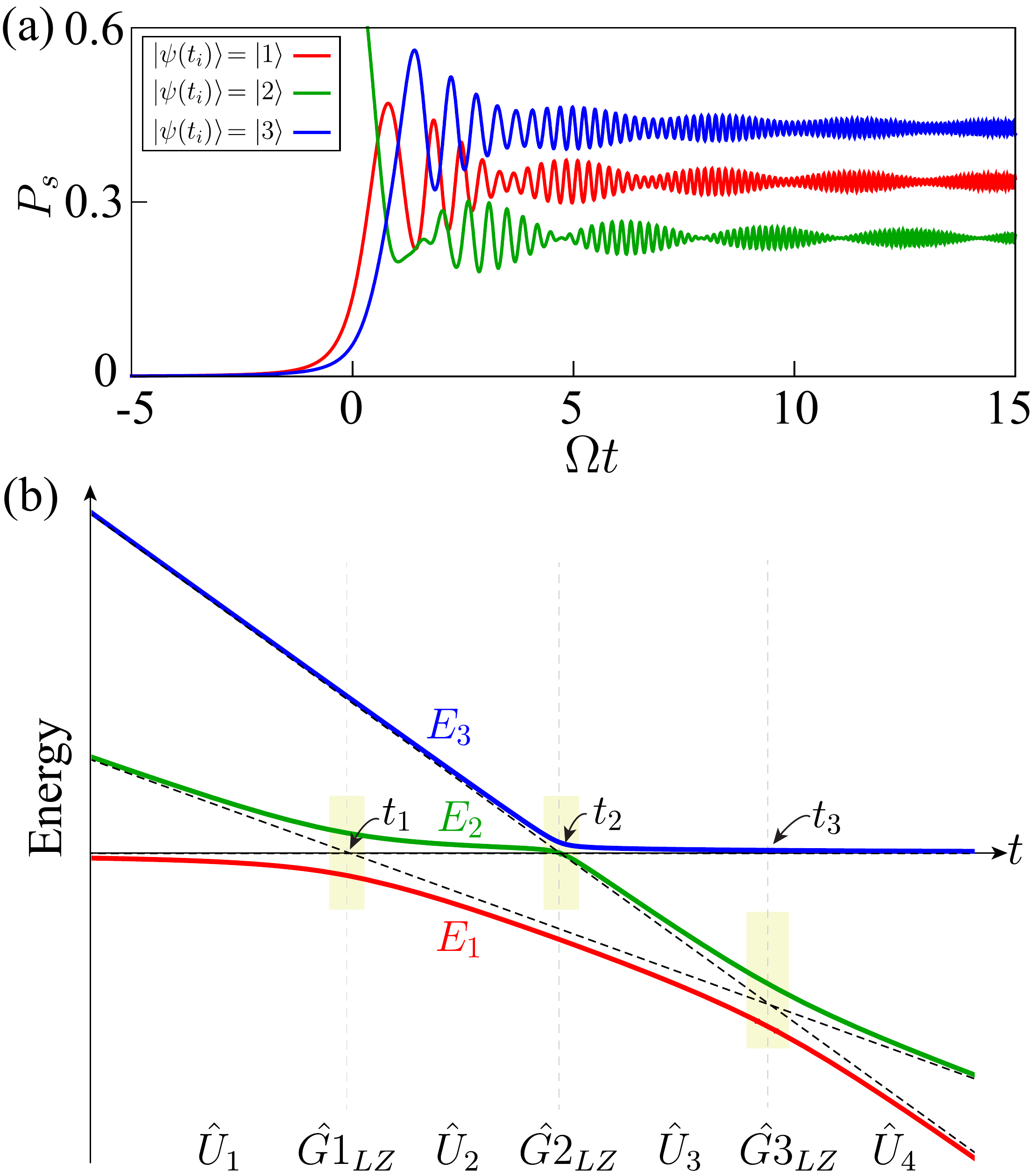}
	\caption{(a) The beats in the dynamics of $P_s(t)$ for $V_0=2\Omega$, $v=5\Omega^2$, and different initial states. (b) The instantaneous energy eigenspectrum for $\Delta(t)=vt$ for large $V_0$. The adiabatic and non-adiabatic  regimes are marked by the operators $\hat U_{1,2,3,4}$ and $\{\hat G1_{LZ}, \hat G2_{LZ},  \hat G3_{LZ}\}$. }
	\label{fig:12}
\end{figure}

\begin{figure}
	\centering
	\includegraphics[width=1.\columnwidth]{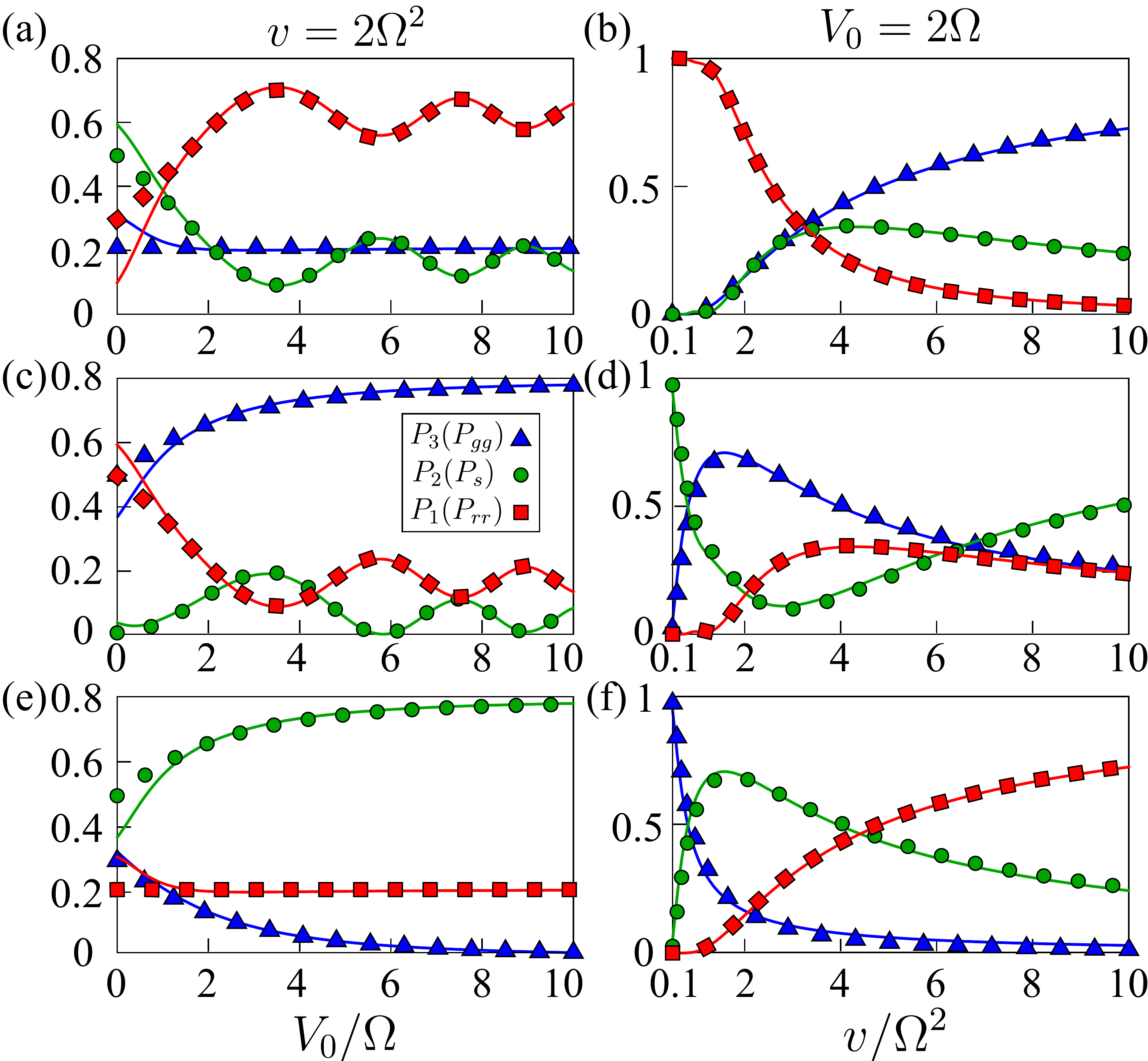}
	\caption{The final population in the adiabatic/diabatic states as a function of both $V_0$ and $v$ for the initial state $|1\rangle$ (a)-(b), $|2\rangle$ (c)-(d), and $|3\rangle$ (e)-(f). For the first column, $v=2\Omega^2$ and for the second column, $V_0=2\Omega$. The solid lines show exact results, and the filled squares, circles, and triangles are from AIA.}
	\label{fig:13}
\end{figure}

Now, we employ AIA for analyzing the dynamics in the three level LZ model in Eq.~(\ref{H3}). To separate adiabatic and non-adiabatic regimes, we require $V_0\gg\Omega$ [see Figs. \ref{fig:4}(b) and \ref{fig:12}(b)]. Further, we assume that only two adiabatic states are involved in each avoided crossings which helps us to use the results from the two-level LZ model discussed in Sec.~\ref{sa}. The validity of AIA requires that the LZT time ($\tau_{LZ}$) to be shorter than the duration ($T_a=V_0/2v$) in which the system evolves adiabatically between two LZTs. Since $\Delta E_0=\Delta E_{V_0}>\Delta E_{V_0/2}$ for $V_0\neq 0$, the upper limit for $\tau_{LZ}$ is set by $\tau_{LZ}\approx (1/2\sqrt{v})\rm{max}(1, \Omega^2/2v)$. Therefore, for $v>\Omega^2/2$, we require $V_0^2>v$ and for $v<\Omega^2/2$,  we require $v<16V_0^2/\Omega^4$ for $AIA$ to be valid. The adiabatic evolution matrix is given by 
\begin{align}
\hat{U}_k
= \left( {
	\begin{array}{ccc} 
	e^{-i\zeta_3^{\{k\}}} & 0 & 0 \nonumber \\
	0 & e^{-i\zeta_2^{\{k\}}} & 0 \nonumber \\
	0 & 0 &  e^{-i\zeta_1^{\{k\}}} 
	\end{array} } \right),
\end{align}
where $\zeta_{j}^{\{1\}} = \int_{t_{i}}^{t_{1}} dtE_{j}$, $\zeta_{j}^{\{2\}} = \int_{t_{1}}^{t_{2}} dtE_{j}$, $\zeta_{j}^{\{3\}} = \int_{t_{2}}^{t_{3}} dtE_{j}$, and $\zeta_{j}^{\{4\}} = \int_{t_{3}}^{t_{f}}dt E_j$  are the phases acquired between the avoided crossings. We define the non-adiabatic transition matrix $\hat G1_{LZ}$ at the impulse point $t_1$ in the basis $\{|3\rangle, |2\rangle, |1\rangle\}$ as,
\begin{equation}
\hat G1_{LZ} = \left( {
	\begin{array}{ccc} 
	1 & 0 & 0 \\
	0&	\sqrt{1-P'_{LZ}}e^{-i\tilde{\phi}'_s} &-\sqrt{P'_{LZ}} \\
	0&	\sqrt{P'_{LZ}} & \sqrt{1-P'_{LZ}}e^{i\tilde{\phi}'_s} \\
	\end{array} } \right)
	\label{g1}
\end{equation}
where $P'_{LZ} = \exp(-2\pi \Omega'^2/4v)$, and
\begin{equation}
\tilde{\phi}'_s = \pi/4 + \text{arg}( \Gamma(1-i \gamma')) + \gamma' (\ln \gamma' -1)
\label{ph1}
\end{equation}
with $\gamma' = \Omega'^2/4v$ and $\Omega'=\Delta E_{0}\sim \sqrt{2}\Omega$. Similarly, the transition matrix at $t_2$ is
\begin{equation}
\hat G2_{LZ} = \left( {
	\begin{array}{ccc} 
	\sqrt{1-P''_{LZ}}e^{-i\tilde{\phi}''_s} &-\sqrt{P''_{LZ}} & 0 \\
	\sqrt{P''_{LZ}} & \sqrt{1-P''_{LZ}}e^{i\tilde{\phi}''_s}  &0 \\
	0 & 0 & 1
	\end{array} } \right)
	\label{g2}
\end{equation}
with $P''_{LZ} = \exp(-2\pi \Omega''^2/8v)$ and
\begin{equation}
\tilde{\phi}_s'' = \pi/4 + \text{arg}( \Gamma(1-i \gamma'')) + \gamma'' (\ln \gamma'' -1)
\end{equation}
with $\gamma'' = \Omega''^2/8v$ and $\Omega''=\Delta E_{V_0/2}$. We have $\hat G1_{LZ}=\hat G3_{LZ}$ since at $t_3$, the LZT involves $|1\rangle$ and $|2\rangle$. The complete evolution matrix in AIA is given by $\hat F_L=\hat{U}_{4} \hat{G3}_{LZ} \hat{U}_{3} \hat{G2}_{LZ} \hat{U}_{2} \hat{G1}_{LZ} \hat{U}_{1}$. The results from AIA are compared to the exact results in Fig.~\ref{fig:13}, for different initial conditions, and as a function of both $v$ and $V_0$. They exhibit good agreement even beyond the criteria discussed above. One reason could be that $\tau_{LZ}$ only sets the upper limit for the transition time, and the actual transition period can be much shorter than that. 

As shown in Figs.~\ref{fig:13}(a) and \ref{fig:13}(c), for a fixed $v$, the final population in states $|s\rangle$ and $|rr\rangle$ exhibit oscillations as a function of $V_0$, indicating the role of quantum interference between the distinct LZTs. On the other hand, for a fixed $V_0$, and varying $v$, we do not observe any oscillations. It indicates that the Stokes phases ($\tilde{\phi}_s'$ and $\tilde{\phi}_s''$) become irrelevant in the final populations if the initial state is one of the instantaneous eigenstates. We have verified this by setting $\tilde{\phi}_s'=\tilde{\phi}_s''=0$ in the matrices $\hat G1_{LZ}$ and $\hat G2_{LZ}$, and the results are hardly affected by it.  If the initial state is not the instantaneous eigenstate, the Stokes phases become important.  In that case, we will be able to observe oscillations in the final populations as a function of $v$ keeping $V_0$ fixed. Ultimately, AIA reveals the different phases involved in the dynamics.

\subsection{Periodic modulation of detuning}
\label{per}

Now, we consider the detuning is varying periodically in time as $\Delta (t) = \Delta_0 +\delta \sin (\omega t)$. We take $V_0\gg\Omega$ and $\Delta_0<< 0$. The first condition assures that the three avoided level crossings are well separated, and we can implement AIA, as discussed in Sec.~\ref{aia2}. The second condition guarantees that the adiabatic states converge to the diabatic ones at the initial time, $t_i=0$. The initial offset in detuning ($\Delta_0$) also play an important role in the dynamics. In the following, we analyze the dynamics for different initial states as a function of $\delta$ and $\omega_0$. The first avoided crossing (involving states $|1\rangle$ and $|2\rangle$) occurs when $\Delta(t)=0$, i.e., at times $\tau^{(1)}_{2n}=[2n\pi+\sin^{-1}\left(-\Delta_0/\delta\right)]/\omega$ and $\tau^{(1)}_{2n+1}=[(2n+1)\pi-\sin^{-1}\left(-\Delta_0/\delta\right)]/\omega$ where $n=0, 1, 2, ...$. Now, linearizing around $\tau^{(1)}_{m}$, i.e., $\Delta(\tau^{(1)}_m+t)=\Delta_0+\delta\sin\omega(\tau^{(1)}_m+t)\approx \delta \omega t\cos\omega\tau^{(1)}_m=(-1)^{m} \omega\sqrt{\delta^2-\Delta_0^2}t$, we obtain the quench rate across the first avoided crossing as $v_1=\pm \omega\sqrt{\delta^2-\Delta_0^2}$. Similarly, the second avoided crossing (involving states $|2\rangle$ and $|3\rangle$) occurs when $\Delta(t)=V_0/2$ or at $\tau^{(2)}_{2n}=[2n\pi+\sin^{-1}\left((V_0/2-\Delta_0)/\delta\right)]/\omega$ and $\tau^{(2)}_{2n+1}=[(2n+1)\pi-\sin^{-1}\left((V_0/2-\Delta_0)/\delta\right)]/\omega$, and the third avoided crossing (again involving states $|1\rangle$ and $|2\rangle$) occurs when $\Delta(t)=V_0$ or at $\tau^{(3)}_{2n}=[2n\pi+\sin^{-1}\left((V_0-\Delta_0)/\delta\right)]/\omega$ and $\tau^{(3)}_{2n+1}=[(2n+1)\pi-\sin^{-1}\left((V_0-\Delta_0)/\delta\right)]/\omega$. The corresponding quench rates are obtained as $v_2=\pm \omega\sqrt{\delta^2-(\Delta_0-V_0/2)^2}$ and $v_3=\pm \omega\sqrt{\delta^2-(\Delta_0-V_0)^2}$, respectively. Appropriately replacing $v$ by $v_1$, $v_2$, and $v_3$, we can use the LZT matrices $\hat G1_{LZ}$, $\hat G2_{LZ}$, and $\hat G3_{LZ}$ to analyze the dynamics via AIA. Using the quench rates, we estimate the upper limit for the LZT time across the avoided crossings at $\tau^{(1)}_{m}$, and $\tau^{(3)}_{m}$ as $\tau_{LZ1} = 1/\sqrt{|v_1|}\max\left(1, \Omega'^2/4|v_1| \right)$ and $\tau_{LZ3} = 1/\sqrt{|v_3|}\max\left(1, \Omega'''^2/4|v_3| \right)$, respectively.  The LZT time for the one at $\tau^{(2)}_{m}$ becomes extremely small (almost instant, as evident from the results shown in Sec. \ref{lzm}) for large $V_0$. 

Note that, the periodic driving results in resonant transitions between different states \cite{bas18}. For instance, in the high-frequency limit ($\omega\gg\Omega$) or the fast-passage limit $\left(\omega\sqrt{\delta^2-\Delta_0^2}\gg\Omega\right)$ a resonant transition between $|gg\rangle$ and $|s\rangle$ takes place when $n\omega=\Delta_0$ with $n=0, \pm 1, \pm 2 ...$. The latter results in coherent Rabi oscillations between the two states. Similarly, for $n\omega=2\Delta_0-V_0$ and $n\omega=\Delta_0-V_0$, we have resonant transition between $|gg\rangle$ and $|rr\rangle$, and $|s\rangle$ and $|rr\rangle$, respectively. To resolve different resonances, we require sufficiently large RRIs. Based on the value of $\delta$, below we consider three cases: (i) $\delta=V_0/4-\Delta_0$, (ii) $\delta=3V_0/4-\Delta_0$, and (iii) $\delta\gg V_0-\Delta_0$. 

\subsubsection{$\delta=V_0/4-\Delta_0$}
\label{pd1}
\begin{figure}
	\centering
	\includegraphics[width=.75\columnwidth]{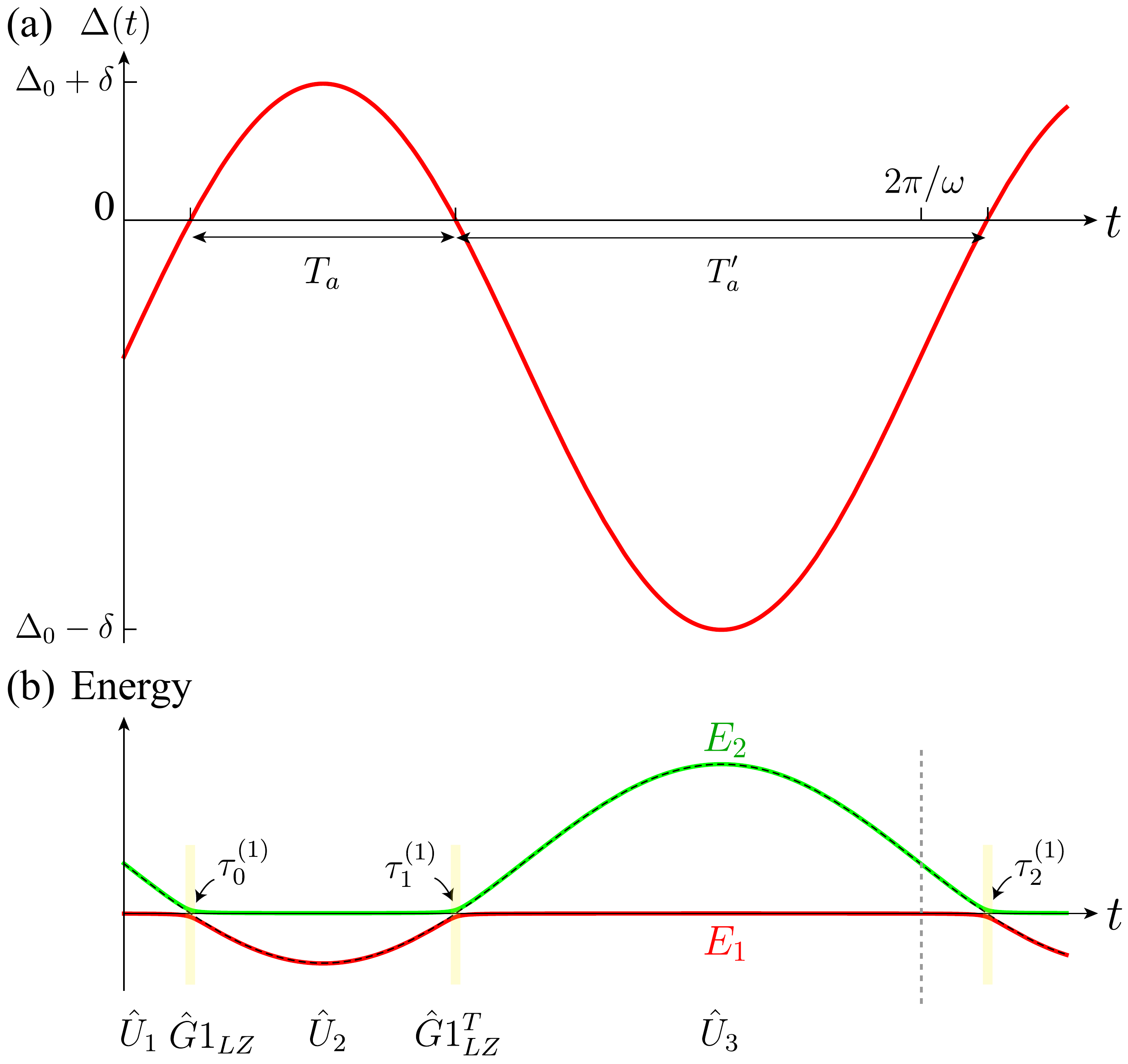}
	\caption{(a) The  periodic time dependence of the detuning for $\delta=V_0/4-\Delta_0$. The expected durations for adiabatic evolution are $T_a$ and $T'_a$. (b) shows the corresponding instantaneous energy eigenvalues. The instants ($\tau_0^{(1)}$, $\tau_1^{(1)}$, $\tau_2^{(1)}$) at which the LZTs occur between states $|1\rangle$ and $|2\rangle$ are shown by shaded stripes. The operators $\hat U_j$  and $\hat G1_{LZ}$ indicate the adiabatic regimes and the impulse points, respectively. Between the origin and the dashed vertical line, we have one complete cycle.}
	\label{fig:14}
\end{figure}
In this case, the detuning varies periodically across the first avoided crossing, and the maximum of $\Delta(t)$ is such that it is in midway between the first and the second avoided crossings. In this case, the state $|3\rangle$ is not part of the LZTs and the evolution matrix for one complete cycle can be written as $\hat F=\hat U_3\hat G1^T_{LZ}\hat U_2\hat G1_{LZ}\hat U_1$ [see Fig.~\ref{fig:14}], and there are three different time scales involved. One is the LZT time $\tau_{LZ1}$ and the two others: $T_a=\tau_1^{(1)}-\tau_0^{(1)}=\left( \pi -2\arcsin(-\Delta_0/\delta)\right)/\omega$ and $T'_a=\tau_2^{(1)}-\tau_1^{(1)}=\left( \pi +2\arcsin(-\Delta_0/\delta)\right)/\omega$ are the adiabatic durations between the two LZTs. We have $T'_a>T_a$ for $\Delta_0<0$, and the validity of AIA requires $\tau_{LZ1}\ll T_a$. Keeping $\delta$, $\Delta_0$ and $V_0$ fixed, and for sufficiently large values of $\omega$, the ratio $\tau_{LZ1}/T_a\propto \sqrt{\omega}$ indicating that AIA might breaks down at large $\omega$.
\begin{figure}
	\centering
	\includegraphics[width=.8\columnwidth]{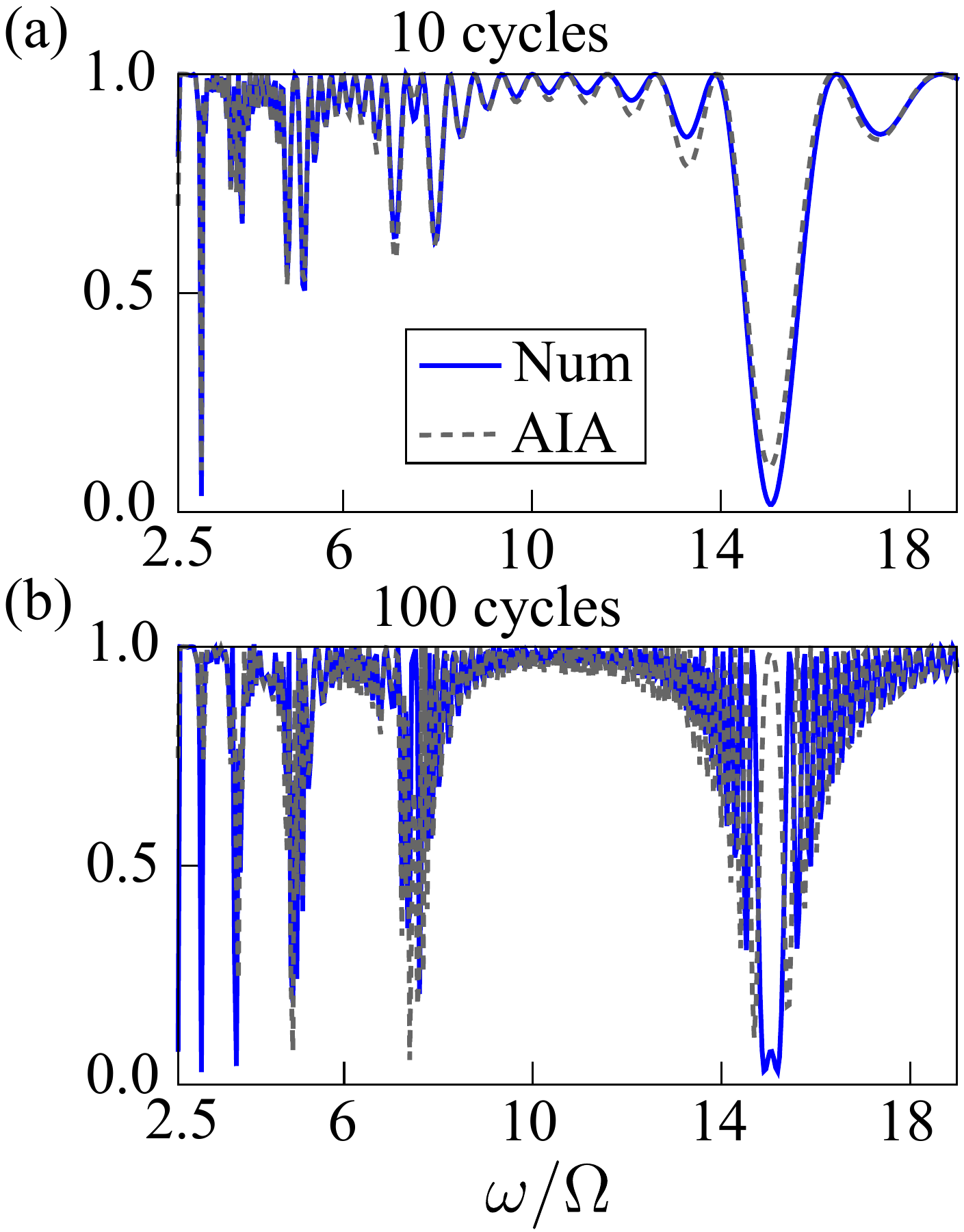}
	\caption{The numerical results (solid lines) and that of AIA (dashed lines) for $P_1$ after (a) 10 and (b) 100 cycles, as a function of $\omega$ for the initial state $|1\rangle\sim|gg\rangle$, $\Delta_0=-15\Omega$, $V_0=40\Omega$ and $\delta=25\Omega$. (b) shows that at longer times, AIA deviates from exact dynamics especially, at high $\omega$.}
	\label{fig:15}
\end{figure}

The final populations in the adiabatic state $|1\rangle$ after 10 and 100 cycles as a function of $\omega$ are shown in Fig.~\ref{fig:15}, for the initial state $|\psi(t=0)\rangle=|1\rangle\sim |gg\rangle$, $\Delta_0=-15\Omega$, $V_0=40\Omega$ and $\delta=25\Omega$. The interference between the LZTs at different times leads to non-trivial oscillations in the populations. Larger the number of cycles, the more non-trivial the pattern is. For shorter periods, the results from AIA are in excellent agreement with the exact ones, whereas at longer periods they start to deviate, which is evident in Fig.~\ref{fig:15}(b). The major dip in  Fig.~\ref{fig:15}(a) is related to the resonance $\omega=|\Delta_0|$. At longer times, there will be significant population in $|3\rangle$ or $|rr\rangle$ but,  the matrix $\hat G1_{LZ}$ does not include the transitions to $|3\rangle$. The latter implies that AIA breaks down in the long time limit. 

\begin{figure}
	\centering
	\includegraphics[width=1.\columnwidth]{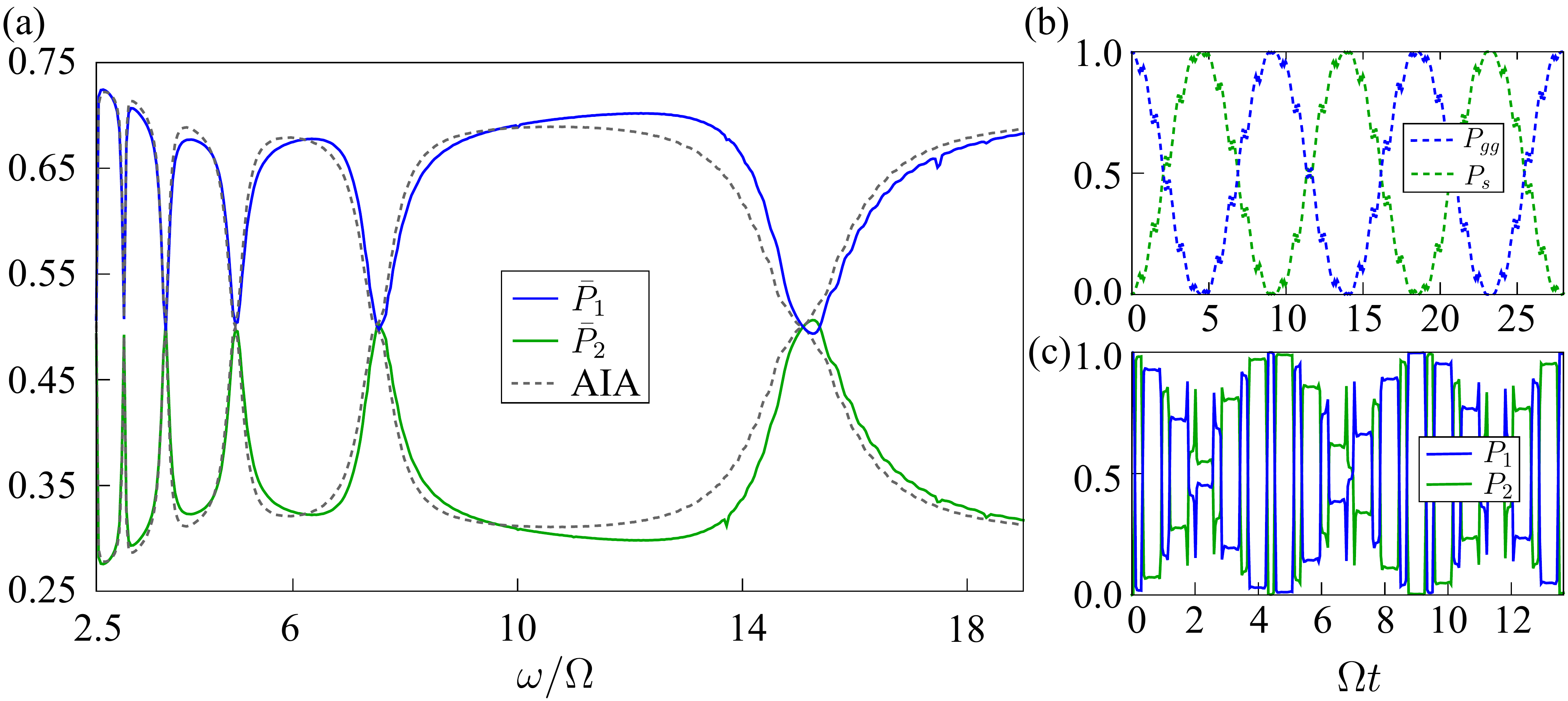}
	\caption{(a) The exact results (solid lines) for $\bar P_{1,2}$, and the same from the AIA (dashed lines) over a period of 100 cycles, as a function of $\omega$ for the initial state $|1\rangle\sim|gg\rangle$, $\Delta_0=-15\Omega$, $V_0=40\Omega$ and $\delta=25\Omega$. The dips (peaks) in $\bar P_1$ ($\bar P_2$) indicate the resonances at $n\omega=|\Delta_0|$. The six resonances are seen at $\omega/\Omega=15, 7.5, 5, 3.75,  3, 2.5$ corresponds to $n=1, 2, 3, 4, 5, 6$, respectively. (b) shows the coherent oscillation between $|gg\rangle$ and $|s\rangle$ at the resonance  $\omega/\Omega=7.5$ and (c) shows the same between $|1\rangle$ and $|2\rangle$ states.}
	\label{fig:16}
\end{figure}

 Fig.~\ref{fig:16}(a) shows the time average populations, $\bar P_{j}=(1/T)\int_0^TP_j(t) dt$ as a function of $\omega$ over a period of 100 cycles with the initial state $|1\rangle$, and other parameters are same as in Fig.~\ref{fig:15}. The resonances at $n\omega=|\Delta_0|$ are seen as dips (peaks) in $\bar P_{1}$ ($\bar P_{2}$). At the resonances, the system exhibit coherent Rabi oscillations between $|gg\rangle$ and $|s\rangle$ or between $|1\rangle$ and $|2\rangle$  [see Figs.~\ref{fig:16}(b) and \ref{fig:16}(c)]. This dynamics is identical to that of two Rydberg atoms under Rydberg blockade with no periodic forcing. Following the similar procedure given in Sec. \ref{aiasa} for the single atom case, we obtain the transition probability to the state $|2\rangle$ after $k$-cycles is 
 \begin{align}
P_2^k &= 4(1-P'_{LZ})P'_{LZ} \sin^2{\phi_s}  \frac{\sin^2 k\alpha}{\sin{\alpha}}
\label{p21}
\end{align}
 where $P'_{LZ} = \exp(-2\pi \Omega'^2/4v_1)$ with $\Omega'=\Delta E_{0}$, and $\cos\alpha={\rm Re}\left((1-P'_{LZ})e^{-i\eta_0} + P'_{LZ}e^{-i\eta_1}\right)$  and $\phi_s=\int_{\tau_0^{(1)}}^{\tau_1^{(1)}}(E_2-E_1)dt/2+\tilde\phi'_s$. The phases $\eta_0$ and $\eta_1$ are a function of dynamical phases acquired during the adiabatic evolution, and $\tilde\phi'_s$ is given by Eq. (\ref{ph1}) but replacing $v$ by $v_1$. Note that Eq.~(\ref{p21}) is identical to Eq.~(\ref{nc2la}) for the single atom case, and hence, all the discussions in Secs.~\ref{aiasa} and \ref{ksl} are valid here. 
 
\begin{figure}
	\centering
	\includegraphics[width=.8\columnwidth]{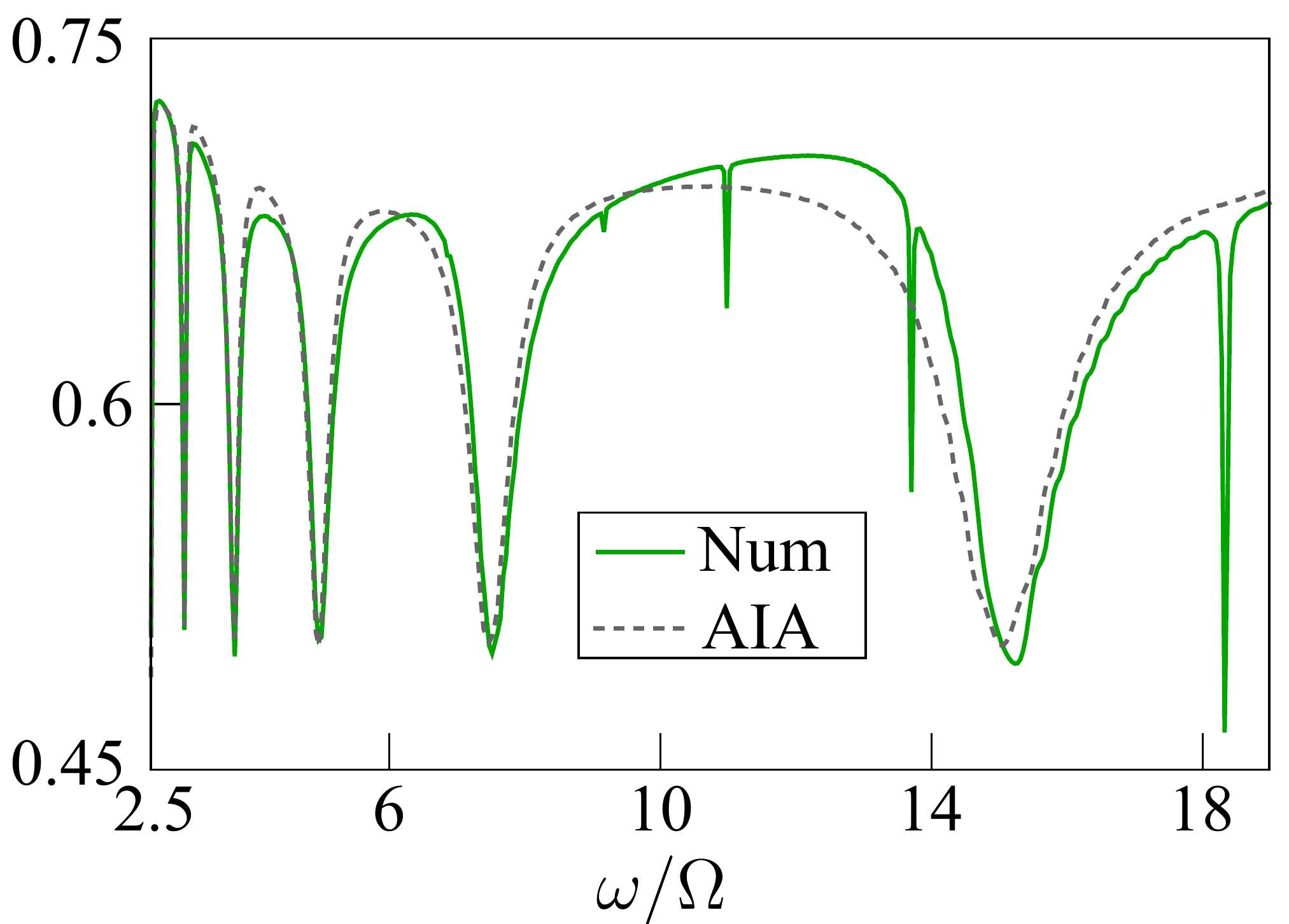}
	\caption{(a) The exact results (solid lines) and that from AIA (dashed lines) of $\bar P_{2}$ over 100 cycles, as a function of $\omega$ for the initial state $|2\rangle\sim|s\rangle$. The dips indicate the resonances, and the five (broader) of them at $\omega/\Omega=15, 7.5, 5, 3.75, 3$ correspond to $n\omega=|\Delta_0|$ with $n=1, 2, 3, 4, 5$, respectively. The narrow ones at $\omega/\Omega=18.33, 13.75, 11$ correspond to $n\omega=|\Delta_0-V_0|$ with $n= 3, 4, 5$, respectively, which are not captured by AIA.}
	\label{fig:17}
\end{figure}

 For the initial state, $|\psi(t=0)\rangle=|2\rangle\sim |s\rangle$ and 10 cycles, the most prominent resonances appear in $\bar P_{1, 2}$ are $n\omega=|\Delta_0|$  (results are not shown). This is similar to that for the initial state $|1\rangle$ except that the role of $|1\rangle$ and $|2\rangle$ are interchanged. For 100 cycles, the resonances, $n\omega=\Delta_0-V_0$, which are much narrower than those at $n\omega=|\Delta_0|$ also emerge in the exact dynamics [see Fig.~\ref{fig:17}]. These narrow resonances at $n\omega=\Delta_0-V_0$ are not captured by AIA. For the initial state $|\psi(t=0)\rangle=|3\rangle\sim |rr\rangle$, AIA completely fails, as the state $|3\rangle$ is not involved in the LZT. The important message from this is that in a multi-state periodically driven system, AIA may not necessarily capture the exact dynamics unless all states are incorporated in the transitions. In other words, considerable modifications in AIA might be required.

\subsubsection{$\delta=3V_0/4-\Delta_0$}

\begin{figure}
	\centering
	\includegraphics[width=.75\columnwidth]{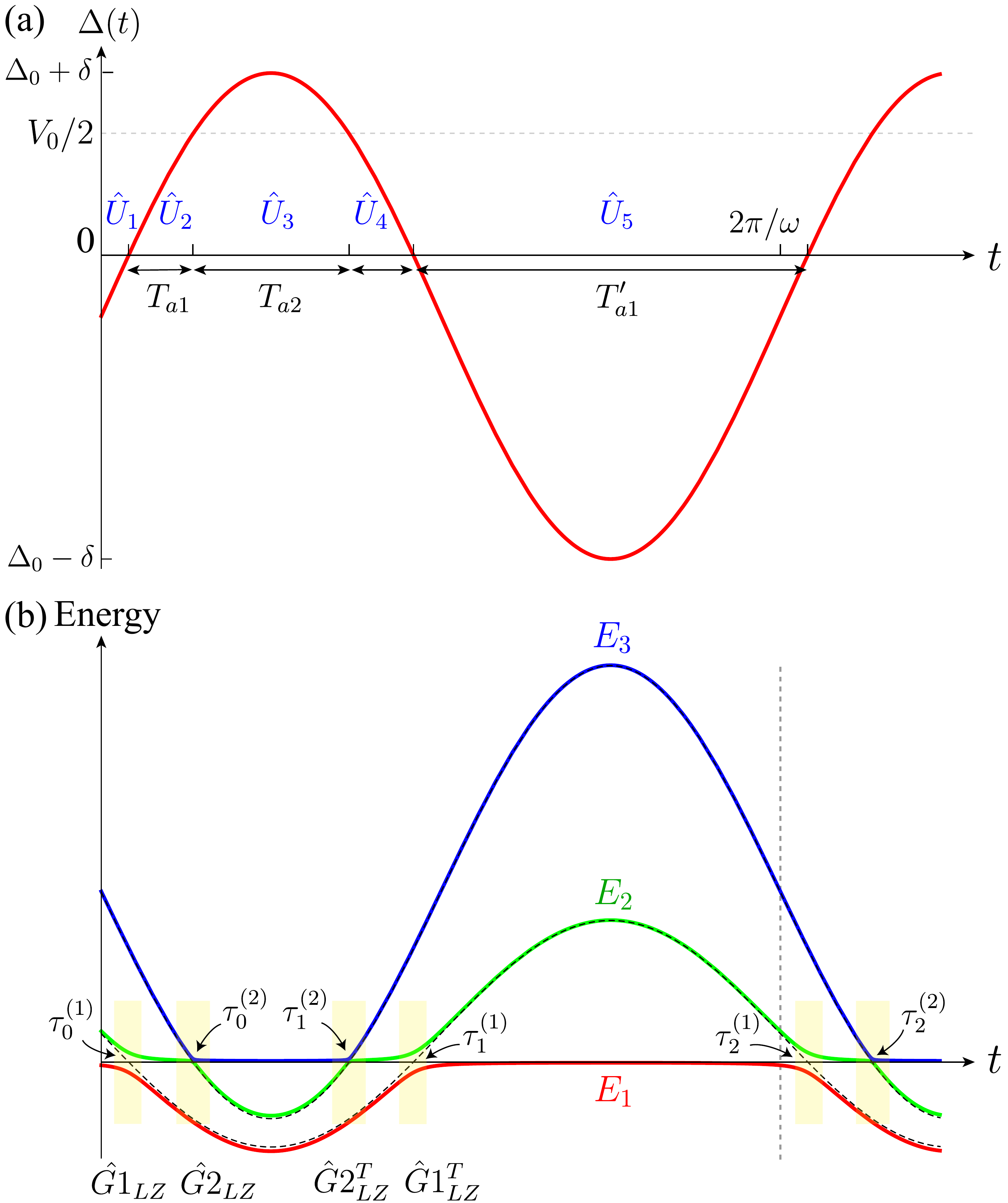}
	\caption{(a) The periodic time dependence of the detuning for $\delta=3V_0/4-\Delta_0$. The adiabatic durations are marked by $T_{a1}$, $T_{a2}$, and $T'_{a1}$. (b) shows the instantaneous energy eigenvalues. The instants ($\tau_0^{(1)}$, $\tau_0^{(2)}$, $\tau_1^{(2)}$, $\tau_1^{(1)}$, $\tau_2^{(1)}$, $\tau_2^{(2)}$) at which the LZTs occur are shown by shaded stripes. The $\hat U$  and $\hat G$ operators indicate the adiabatic regimes and the impulse points, respectively. Between the origin and the dashed vertical line, we have one complete cycle.}
	\label{fig:18}
\end{figure}

Now, we periodically drive across the first two avoided crossings, and the maximum of $\Delta(t)$ comes in between the second and the third avoided crossings [see Figs.~\ref{fig:12}(b) and \ref{fig:18}].  In contrast to the previous case, here, all three adiabatic states are involved in the LZTs but the diabatic state $|rr\rangle$ is excluded. There are two LZT times: $\tau_{LZ1}$ and $\tau_{LZ2}$ for the transitions at $\Delta(t)=0$ and $\Delta(t)=V_0/2$, respectively. The LZTs at  $\tau_{2n}^{(1)}$ and $\tau_{2n}^{(2)}$ are characterized by the transition matrices $\hat G1_{LZ}$ and $\hat G2_{LZ}$ with $v$ being replaced by $v_1$ and $v_2$, respectively. There are four different adiabatic intervals [see Fig.~\ref{fig:18}], and as far as the validity of AIA is concerned, only the shortest among them matters. Once we fix $\delta=3V_0/4-\Delta_0$, the shortest adiabatic duration is given by $T_a=\tau_0^{(2)}-\tau_0^{(1)}= \left[ \sin^{-1}(-[\Delta_0-V_0/2]/\delta) - \sin^{-1}(-\Delta_0/\delta) \right]/\omega$, and the validity of AIA requires $T_a\gg \tau_{LZ1}, \tau_{LZ2}$. Again, the latter implies that for large values $\omega$, the AIA might breaks down. With two avoided crossings, the evolution matrix for one complete cycle [see Fig.~\ref{fig:18}(b)] becomes $\hat{F}= \hat{U}_{5} \hat{G1}_{LZ}^T \hat{U}_{4} \hat{G2}_{LZ}^T \hat{U}_{3} \hat{G2}_{LZ} \hat{U}_{2} \hat{G1}_{LZ} \hat{U}_{1}$ and for $k$-cycles it is $\hat F^k$.

\begin{figure}
	\centering
	\includegraphics[width=.8\columnwidth]{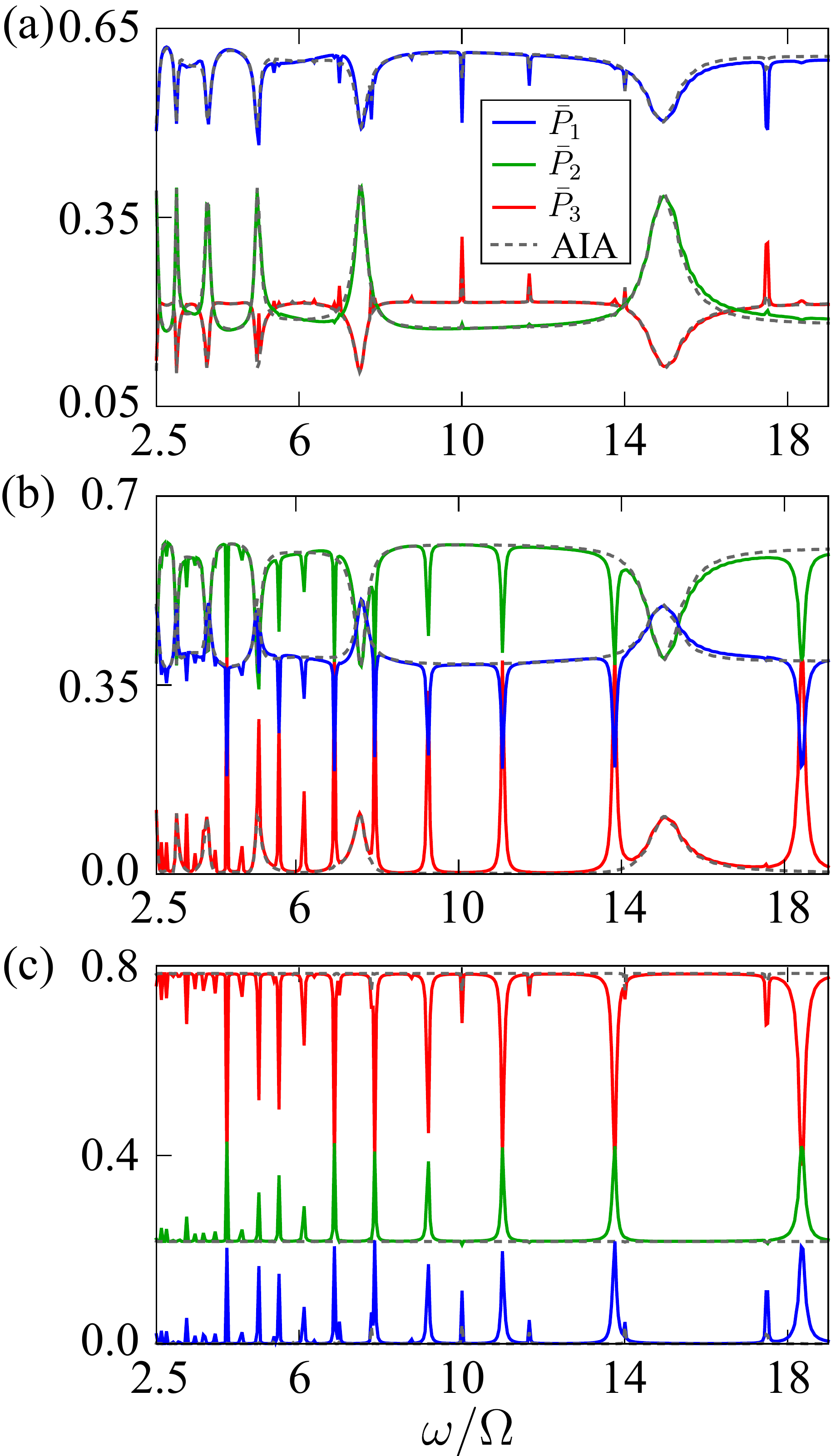}
	\caption{The exact results (solid lines) for the time-averaged populations ($\bar P_{1,2,3}$) and the same from the AIA (dashed lines) for a period of 100 cycles, as a function of $\omega$ for $\Delta_0=-15\Omega$, $V_0=40\Omega$ and $\delta=45\Omega$ with the initial state (a) $|1\rangle\sim |gg\rangle$, (b) $|2\rangle\sim|s\rangle$ and (c) $|3\rangle\sim|rr\rangle$. In (c) AIA completely failed to capture any resonances.}
	\label{fig:19}
\end{figure}

 Fig.~\ref{fig:19} shows the time average populations in the adiabatic states as a function of $\omega$ over a period of 100 cycles for $\Delta_0=-15\Omega$, $V_0=40\Omega$ and $\delta=45\Omega$, and different initial states. The solid lines show the exact results and the dashed ones are from AIA. For the initial state $|\psi(t=0)\rangle=|1\rangle\sim |gg\rangle$ [see  Fig. \ref{fig:19}(a)], we see the resonances $n\omega=|\Delta_0|$ and $n\omega=|2\Delta_0-V_0|$. Note that, the latter resonances (for e.g. $\omega/\Omega=17.5$) which correspond to the resonant transition between $|gg\rangle$ and $|rr\rangle$ (anti-blockades) via $|s\rangle$ is not captured by AIA since $|rr\rangle$ is not included. Figs.~\ref{fig:19}(b) and \ref{fig:19}(c) show the average populations for the initial states $|\psi(t=0)\rangle=|2\rangle\sim |s\rangle$ and $|\psi(t=0)\rangle=|3\rangle\sim |rr\rangle$, respectively. We see more resonances in these cases. For the initial state $|s\rangle$, the resonances correspond to the transition between $|s\rangle$ and $|gg\rangle$ ($n\omega=|\Delta_0|$), and $|s\rangle$ and $|rr\rangle$ ($n\omega=|\Delta_0-V_0|$) can be seen. On the other hand, with the initial state $|rr\rangle$, we have the resonances associated with $|rr\rangle\leftrightarrow |s\rangle$, and $|rr\rangle\leftrightarrow|gg\rangle$ transitions. In all these cases, AIA failed to captured any resonances which involves $|rr\rangle$, and therefore no resonant features are observed for the initial state $|rr\rangle$ as seen in Fig. \ref{fig:19}(c). 
   
\subsubsection{$\delta=5V_0/4-\Delta_0$}
\begin{figure}
	\centering
	\includegraphics[width=.75\columnwidth]{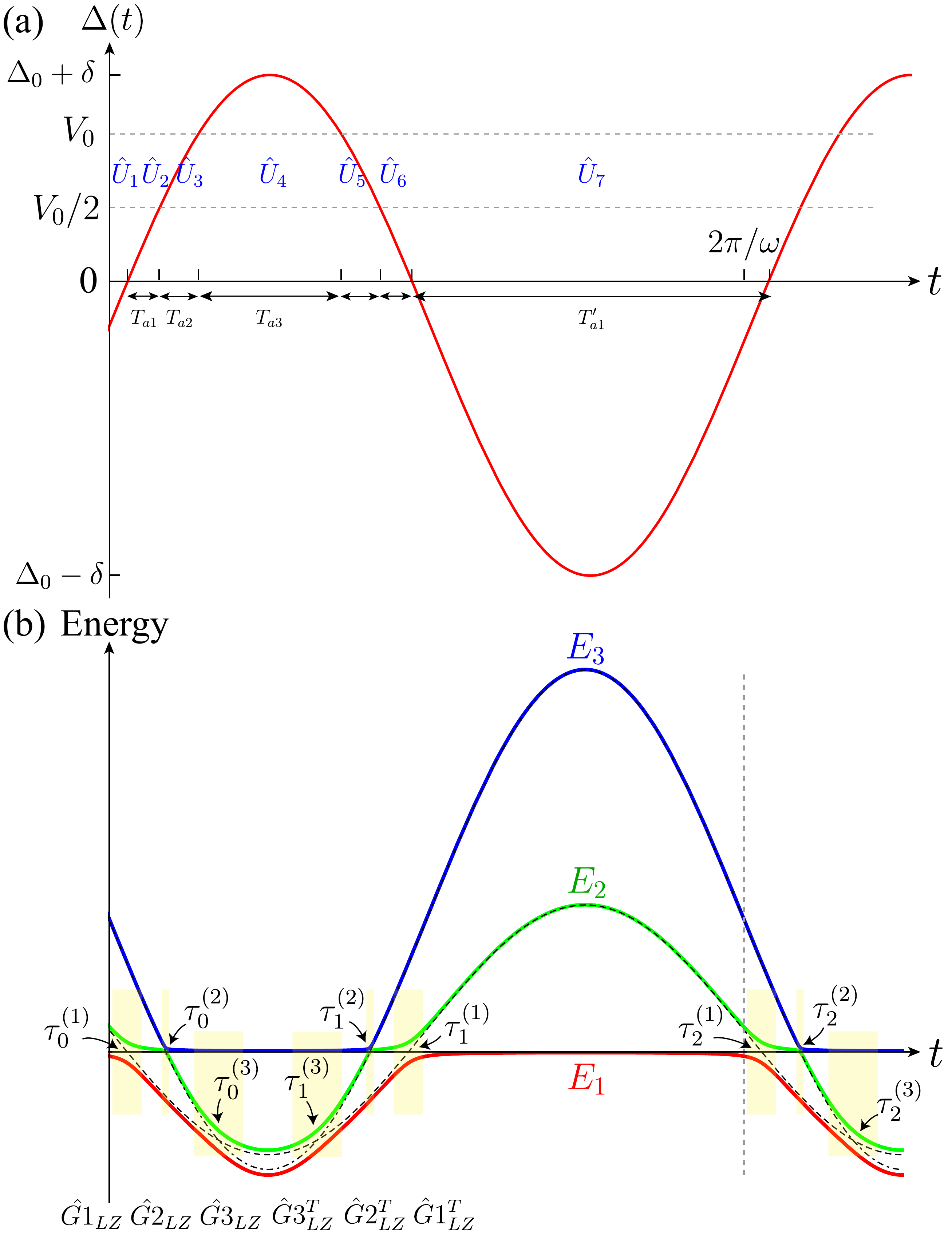}
	\caption{(a) The  periodic time dependence of the detuning for $\delta=5V_0/4-\Delta_0$ and (b) shows the instantaneous energy eigenvalues. The adiabatic durations are marked by $T_{a1}$, $T_{a2}$, $T_{a3}$, and $T'_{a1}$. The instants ($\tau_0^{(1)}$, $\tau_0^{(2)}$, $\tau_0^{(3)}$, $\tau_1^{(3)}$, $\tau_1^{(2)}$, $\tau_1^{(1)}$, $\tau_2^{(1)}$, $\tau_2^{(2)}$, $\tau_2^{(3)}$) at which the LZTs occur between different adiabatic states are shown by shaded stripes. The $\hat U$  and $\hat G$ operators represent the adiabatic regions and impulse points, respectively. Between the origin and the dashed vertical line, we have one complete cycle.}
	\label{fig:20}
\end{figure}

\begin{figure}
	\centering
	\includegraphics[width=.8\columnwidth]{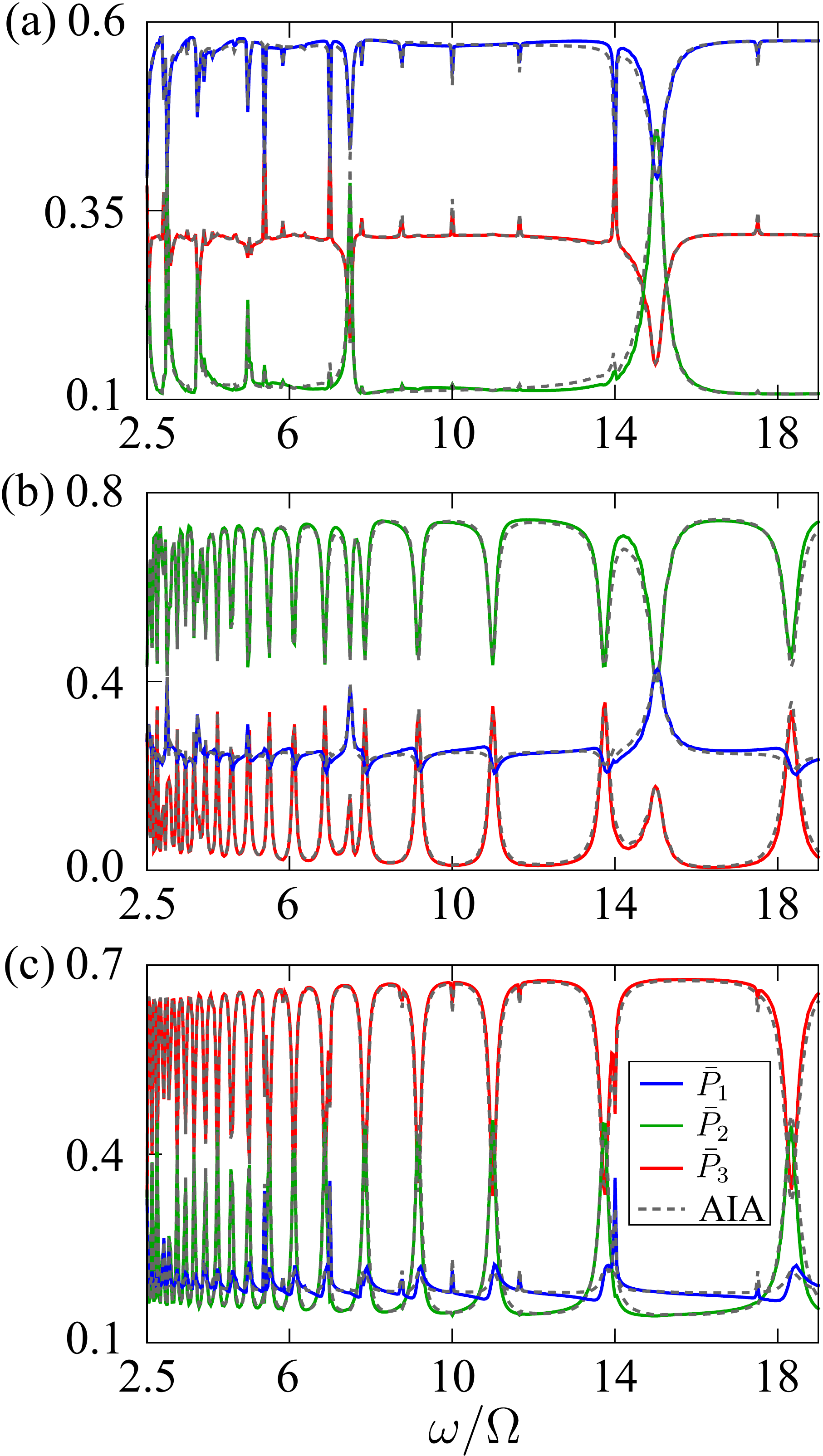}
	\caption{The numerical results (solid lines) for the time-averaged populations ($\bar P_{1,2,3}$) in the adiabatic states, and the same from the AIA (dashed lines) over a period of 100 cycles, as a function of $\omega$ for the initial state (a) $|1\rangle$, (b) $|2\rangle$ and (c) $|3\rangle$. Other parameters are $\Delta_0=-15\Omega$, $V_0=40\Omega$ and $\delta=65\Omega$. In (a) the major peaks correspond to the resonances $n\omega=|\Delta_0|$ (between $|gg\rangle$ and $|s\rangle$ states) and smaller ones indicate the resonances at $n\omega=|\Delta_0-V_0|$ (between $|rr\rangle$ and $|s\rangle$ states). In (b) the peaks/dips indicate the resonances at $n\omega=|\Delta_0|$ and $n\omega=|\Delta_0-V_0|$, with no traces on the resonances at $n\omega=|2\Delta_0-V_0|$. In (c) except the resonances at $n\omega=|\Delta_0|$, other two types are seen. AIA results are in excellent agreement with the numerics in all three cases.}
	\label{fig:21}
\end{figure}

For the last case, the modulation amplitude is such that the system is periodically driven across all three avoided crossings. Therefore, all three adiabatic and diabatic states are involved in LZTs and hence, in AIA. There are three LZT times involved in the dynamics: $\tau_{LZ1}$, $\tau_{LZ2}$, and $\tau_{LZ3}$ for the transitions at $\Delta(t)=0$, $\Delta(t)=V_0/2$ and $\Delta(t)=V_0$, respectively, and  we have $\tau_{LZ2}\ll \tau_{LZ1}, \tau_{LZ3}$. As far as the validity of AIA is concerned, the shortest duration of adiabatic evolution [$T_{a1}$ in Fig. \ref{fig:20}(a)] should be larger than both $\tau_{LZ1}$ and $\tau_{LZ3}$. The evolution matrix for one complete cycle in AIA [see Fig.~\ref{fig:20}] is $\hat{F}= \hat{U}_{7} \hat{G1}_{LZ}^T \hat{U}_{6} \hat{G2}_{LZ}^T  \hat{U}_{5} \hat{G3}_{LZ}^T \hat{U}_{4} \hat{G3}_{LZ} \hat{U}_{3} \hat{G2}_{LZ} \hat{U}_{2} \hat{G1}_{LZ} \hat{U}_{1}$ and for $k$-cycles, it is $\hat F^k$. The operators, $\hat U$s are the adiabatic evolution matrices, and the LZT matrices $\hat{G1}_{LZ}$ and $\hat{G2}_{LZ}$ are provided by Eqs.~(\ref{g1}) and (\ref{g2}), with $v$ being replaced by $v_1$ and $v_2$, respectively. The third LZT matrix is,
 \begin{equation}
\hat{G3}_{LZ} = \left( {
	\begin{array}{ccc} 
	1 & 0 & 0 \\
	0&	\sqrt{1-P_{LZ}'''}e^{-i\tilde{\phi}_{s3}} &-\sqrt{P_{LZ}'''} \nonumber \\
	0&	\sqrt{P_{LZ}'''} & \sqrt{1-P_{LZ}'''}e^{i\tilde{\phi}_{s3}} \nonumber\\
	\end{array} } \right)
\end{equation}
where $P_{LZ}''' = \exp(-2\pi \Omega'''^2/4v_3)$ with $\Omega'''=\Delta E_{V_0}$,  
$\tilde{\phi}_{s3} = \frac{\pi}{4} + \text{arg}( \Gamma(1-i \gamma''')) + \gamma''' (\ln \gamma''' -1)$ with $\gamma''' = \Omega'''^2/4v_3$ and $v_3 = \omega\sqrt{\delta^2 -(\Delta_0-V_0)^2}$.

 In Fig. \ref{fig:21} we show the time average populations in the adiabatic states as a function of $\omega$ over a period of 100 cycles, $\Delta_0=-15\Omega$, $V_0=40\Omega$, $\delta=65\Omega$, and for all three initial states. In Fig. \ref{fig:20}(a), for the initial state $|\psi(t=0)\rangle=|1\rangle\sim |gg\rangle$, we observe major peaks corresponding to the resonances $n\omega=|\Delta_0|$, and minor ones for the resonances at $n\omega=|2\Delta_0-V_0|$. For the initial state $|\psi(t=0)\rangle=|2\rangle\sim |s\rangle$ [see Fig. \ref{fig:21}(b)] we observe resonances at $n\omega=|\Delta_0|$ and $n\omega=|\Delta_0-V_0|$. Finally, for the initial state $|\psi(t=0)\rangle=|3\rangle\sim |rr\rangle$, except the resonances at $n\omega=|\Delta_0|$, other two types are captured. In contrary to the previous two cases, here, AIA is able to capture all possible resonant transitions. Thus, from the above examples, we conclude that, for AIA to be successful in a periodically driven multi-level system especially, at longer times, it is necessary to incorporate the transition matrices across all avoided crossings. Once successful, AIA reveals to us the web of phases involved in the dynamics which can find applications in developing quantum technologies.


\section{Summary and outlook}
\label{summ}
In summary, we analyzed the LZ dynamics in a setup of two Rydberg-atoms with a time-dependent detuning, both linear and periodic. As we have shown, the Rydberg-atom setup realizes different LZ models, for instance, the bow-tie model and the triangular LZ model. Since state of the art Rydberg setups deal with strong RRIs, the triangular LZ model can be tested in these systems through chirping the frequency of laser field, which couples the ground to the Rydberg state \cite{mal01, con02,lam02, mae11,ber17}. The periodically driven Rydberg setup, for instance, can be realized by frequency modulation \cite{noe98}. We identified a striking similarity with the excitation probability in a single periodically driven two-level atom to the intensity distribution from a narrow antenna array. For two atoms (which can be easily realizable using optical tweezers or microscopic optical traps \cite{beg13}), the LZ dynamics showed a non-trivial dependence on the initial state, the quench rate, and the interaction strength. We discussed in detail the validity of AIA in describing the dynamics for both linear and periodic variation of detuning. Interestingly, AIA reveals detailed information about the phases developed during the dynamics, which can be very useful for applications such as coherent control of quantum states, implementing quantum (phase) gates \cite{hua18, wu19}, and atom-interferometry \cite{sil06}.

While implementing AIA, we rely on large RRIs for which the LZTs across each avoided crossings involve only two adiabatic states. For small interactions, it is required to develop a multi-level AIA in which the LZTs take place among multiple levels at the same time. Our study can be extended to three two-level atoms, for which it will not be so straight forward to assume AIA would work at large interactions due to the complexity in the level structure.

\section{Acknowledgments}
We acknowledge UKIERI- UGC Thematic Partnership No. IND/CONT/G/16-17/73 UKIERI-UGC project, UGC for UGC-CSIR NET-JRF/SRF, the support from the EPSRC through Grant No. EP/M014266/1 and Grant No. EP/R04340X/1 via the QuantERA project ERyQSenS.

\bibliographystyle{apsrev4-1}
\bibliography{lib.bib}
\end{document}